\documentclass[english,aps,jcp,reprint,superscriptaddress]{revtex4-1}
\usepackage[T1]{fontenc}
\usepackage[latin9]{inputenc}
\usepackage{geometry}
\usepackage{float}
\usepackage{amsmath}
\usepackage{amssymb}
\usepackage{graphicx}

\makeatletter

\providecommand{\tabularnewline}{\\}

\usepackage{babel}

\allowdisplaybreaks

\makeatother

\usepackage{babel}
\begin{document}
\title{Triplet harvesting in the polaritonic regime: a variational polaron
approach}
\author{Luis A. Martínez-Martínez}
\affiliation{Department of Chemistry and Biochemistry, University of California
San Diego, La Jolla, California 92093, United States}
\author{Elad Eizner}
\affiliation{Department of Engineering Physics, École Polytechnique de Montréal,
Montréal H3C 3A7, QC, Canada}
\author{Stéphane Kéna-Cohen}
\affiliation{Department of Engineering Physics, École Polytechnique de Montréal,
Montréal H3C 3A7, QC, Canada}
\author{Joel Yuen-Zhou}
\affiliation{Department of Chemistry and Biochemistry, University of California
San Diego, La Jolla, California 92093, United States}
\begin{abstract}
We explore the electroluminescence efficiency for a quantum mechanical
model of a large number of molecular emitters embedded in an optical
microcavity. We characterize the circumstances under which a microcavity
enhances harvesting of triplet excitons via reverse intersystem-crossing
(R-ISC) into singlet populations that can emit light. For that end,
we develop a time-local master equation in a variationally optimized
frame which allows for the exploration of the population dynamics
of chemically relevant species in different regimes of emitter coupling
to the condensed phase vibrational bath and to the microcavity photonic
mode. For a vibrational bath that equilibrates faster than R-ISC (in
emitters with weak singlet-triplet mixing), our results reveal that
significant improvements in efficiencies with respect to the cavity-free
counterpart can be obtained for strong coupling of the singlet exciton
to a photonic mode, as long as the singlet to triplet exciton transition
is within the inverted Marcus regime; under these circumstances, the
activation energy barrier from the triplet to the lower polariton
can be greatly reduced with respect to that from the triplet to the
singlet exciton, thus overcoming the detrimental delocalization of
the polariton states across a macroscopic number of molecules. On
the other hand, for a vibrational bath that equilibrates slower than
R-ISC (\emph{i.e.}, emitters with strong singlet-triplet mixing),
we find that while enhancemnents in photoluminiscence can be obtained
via vibrational relaxation into polaritons, this only occurs for small
number of emitters coupled to the photon mode, with delocalization
of the polaritons across many emitters eventually being detrimental
to electroluminescence efficiency. These findings provide insight
on the tunability of optoelectronic processes in molecular materials
due to weak and strong light-matter coupling.
\end{abstract}
\pacs{Strong light-matter coupling, dark states, microcavity, polaron transformation,
variational methods }

\maketitle
\setlength{\abovedisplayskip}{3pt} \setlength{\belowdisplayskip}{3pt}

\section{Introduction\label{sec:Introduction}}

The strong light-matter coupling (SC) or polaritonic regime reached
with organic molecules embedded in confined electromagnetic environments,
such as optical microcavities, has attracted considerable attention
due to the possibilities it offers to the control of chemical processes
at the molecular level \cite{Torma2014,Ebbesen2016_rev,Ribeiro2018_rev,Feist2018_rev,Flick2018_rev}.
In this regime, the interaction energy between the molecular ensemble
and the photonic modes surpasses their respective linewidths, resulting
in the formation of hybrid photon-matter excitations termed polaritons
\cite{Hopfield1958}. The novel tunability of chemical dynamics and
related processes afforded in the SC regime with organic molecules
has been proven in photoisomerization \cite{Hutchison2012}, excitation
energy transfer \cite{Zhong2016,Du2018,Saez2018}, optical selection
rules \cite{Neuman2018}, exciton transport \cite{Feist2015,Schachen2015}
and more recently triplet harvesting \cite{Kena2007,stranius2018,Polak2018}.
The SC regime has not been exclusively harnessed at optical frequencies,
as unique signatures of chemical control \cite{Anoop2016,Dunkelberger2016,Chervy2017}
and nonlinear responses \cite{Ribeiro2018_nonlin,Xiang2018} have
also been demonstrated for molecular vibrational transitions coupled
to a microcavity field.

Polariton setups have also emerged as promising candidates to boost
the efficiency and versatility of light-emitting diodes (LEDs) \cite{Bajoni2018}
and organic photodiodes (OPDs) \cite{Eizner2018}. As a proof of concept,
electrical injection of carriers into inorganic polariton architectures
has been shown to modulate light-matter coupling \cite{Chakraborty2018},
thus opening new avenues towards polariton-based optoelectronic switches.
Similarly, organic LEDs have been demonstrated utilizing molecular
dyes in the ultrastrong coupling regime \cite{Gubbin2014,Mazzeo2014,Genco2018}
and it has been recently shown that materials operating in the latter
can feature a complete inversion of molecular dark and light-emissive
states \cite{Eizner2019}. These exciting applications prompt the
development of theoretical models to account for experimental observations
of the molecular SC regime \cite{delPino2015,Galego2016,Herrera2016,Herrera2017},
as well as to rationally design new polaritonic setups that could
enhance chemical processes of contemporary interest \cite{Galego2017,Martinez2018}.

In this article, we explore the triplet electroluminescence efficiency
of a microcavity containing molecules which feature a range of electronic
parameters and couplings to the condensed phase vibrational bath.
Our model aims to describe polaritonic OLEDs like those reported in
\cite{Tischler2005,Genco2018}, so we consider that (optically dark)
triplet excitons are generated upon electrical injection and the latter
can transition into fluorescent singlet states that emit light. Our
approach relies on a master equation operating in a variationally
optimized polaron frame, originally introduced by Silbey and Harris
\cite{Silbey1984,Harris1985} with generalizations due to Pollock
\cite{Pollock2013}, Wu \cite{Wu2016}, and their respective coworkers.
The foundation of this method lies on the application of a unitary
transformation to the total Hamiltonian which yields renormalized
system and system-bath interactions that are weak enough to be perturbatively
treated.

The outline of this article is summarized as follows. In section \ref{sec:Theory},
we introduce our quantum mechanical model and the variational approach
to address its dynamics. Next, in section \ref{sec:Dynamics-in-the},
we give a formal definition of the triplet electroluminescence efficiency
and describe our approach to calculate the time-evolution of populations
in the relevant chemical states of the system. Then, in section \ref{sec:Results-and-discussion},
we apply our approach to systems in two regimes of coupling to vibrational
degrees of freedom. In section \ref{sec:largeN}, we identify the
main limitation that must be overcome to reach a polariton-enhanced
triplet-harvesting regime and propose an approach to circumvent this
drawback. Finally, in section \ref{sec:Conclusions} we summarize
our study and conclude with an outlook of how polaritons could enhance
the optoelectronic properties of molecular materials.

\section{Theory\label{sec:Theory}}

We consider an ensemble of $N$ identical molecules embedded in an
optical microcavity and interacting with the electromagnetic modes
supported by the latter; for simplicity, we describe a single photonic
mode interacting with the molecular ensemble, a coarse-graining approximation
which is based on the much larger density of states (DOS) of the molecular
degrees of freedom compared to the photonic ones \cite{delPino2015,daskalakis2017}.
Thus, $N$ should not be interpreted as the total number of molecules
in the cavity, but rather as the average number of molecules that
couple to a single photon mode.

The Hamiltonian of our model can be written as 
\begin{align}
H= & H_{S}+H_{B}+H_{S-B}+H_{ph}\nonumber \\
 & +H_{S-ph}+H_{T}+H_{T-B}+H_{S-T}.\label{eq:Hamiltonian}
\end{align}
Each molecule is modeled as a three-level electronic system, namely
a singlet electronic ground state and two excited states with singlet
and triplet spin characters, respectively. The energetics of the latter
are correspondingly described by $H_{S}$ and $H_{T}$. Assuming electrical
pumping in the linear regime, we can restrict the Hamiltonian to the
single excitation manifold such that, 

\begin{subequations}
\begin{align}
H_{S} & =\sum_{n=0}^{N-1}\epsilon_{S}|n\rangle\langle n|,\label{eq:H_S}\\
H_{T} & =\sum_{n=0}^{N-1}\epsilon_{T}|T_{n}\rangle\langle T_{n}|.\label{eq:H_T}
\end{align}

\end{subequations}\noindent Here, $|n\rangle$ ($|T_{n}\rangle$)
denotes a localized singlet (triplet) exciton at the $n$th molecular
site and $\epsilon_{S}$ ($\epsilon_{T}$) is the vertical singlet
(triplet) electronic excitation energy, while ($\hbar=1$)

\begin{equation}
H_{B}=\sum_{v}\sum_{n=0}^{N-1}\omega_{v}b_{v,n}^{\dagger}b_{v,n}\label{eq:H_B}
\end{equation}
accounts for the vibrational degrees of freedom in the condensed phase
environment, where $b_{v,n}^{\dagger}$ ($b_{v,n}$) is the creation
(annihilation) operator for an excitation in the $v$-th harmonic
mode with frequency $\omega_{v}$ on site $n$. The mode-dependent
singlet (triplet) vibronic couplings $g_{S}^{(v)}$ ($g_{T}^{(v)}$)
are included in $H_{S-B}$ ($H_{T-B}$):

\begin{subequations}

\begin{align}
H_{S-B} & =\sum_{v}\sum_{n}|n\rangle\langle n|g_{S}^{(v)}(b_{v,n}+\text{h.c.}),\label{eq:H_SB}\\
H_{T-B} & =\sum_{v}\sum_{n}|T_{n}\rangle\langle T_{n}|g_{T}^{(v)}(b_{v,n}+\text{h.c.}).\label{eq:H_TB}
\end{align}

\end{subequations}\noindent The singlet-triplet intersystem crossing
electronic coupling is given by 
\begin{equation}
H_{S-T}=V_{ST}\sum_{n}\left(|n\rangle\langle T_{n}|+\text{h.c.}\right).\label{eq:H_ST}
\end{equation}
Finally, the photonic microcavity degree of freedom is encoded in
\begin{equation}
H_{ph}=\omega_{ph}|G,1_{ph}\rangle\langle G,1_{ph}|,\label{eq:H_ph}
\end{equation}
where $|G,1_{ph}\rangle$ accounts for the state of all molecules
in the electronic ground state and one excitation in the photonic
mode and its interaction with singlet excitons (in the rotating wave
approximation, RWA) is described by 
\begin{align}
H_{S-ph} & =\sum_{n}g\left(|G,1_{ph}\rangle\langle n|+\text{h.c.}\right),\label{eq:H_S_ph}
\end{align}
$g$ being a single-molecule dipolar coupling. To describe the emergent
dynamics upon electrical pumping of the optically dark triplet states,
we employ a variational polaron transformed master equation. The latter
has proven to reproduce reliable quantum dynamics in a wide range
of coupling strengths between electronic degrees of freedom and phononic
reservoirs \cite{Lee2012}, while being computationally inexpensive
compared to more accurate approaches such as path integral \cite{Makri1995,Makri1995II},
multi-configuration time-dependent Hartree \cite{Meyer1990,Beck2000},
hierarchical equations of motion \cite{Ishizaki2009}, linearized
density matrix evolution \cite{Huo2011}, surface-hopping \cite{Tully1990,Prezhdo1997,Subotnik2016}
and exact factorization \cite{Hoffmann2018} formulations. In particular,
we consider a similar multi-site approach to the one developed by
Pollock and coworkers \cite{Pollock2013}, who considered a variational
polaron transformation to compute population dynamics in exciton networks
with local phonon reservoirs. Furthermore, we adapt a generalization
introduced by Wu and coworkers \cite{Wu2016} who treated the photon-exciton
and exciton-vibration on equal footing to describe the photoluminiscence
and vibrational dressing in polaritonic setups. As a first step towards
the development of a master equation, the Hamiltonian in Eq. (\ref{eq:Hamiltonian})
is transformed to the so-called polaron frame, for which we introduce
the unitary transformation $e^{P}$, where

\begin{align}
P= & \sum_{n=0}^{N-1}\sum_{v}|n\rangle\langle n|\left[\sum_{l}(b_{v,n+l}^{\dagger}-b_{v,n+l})\frac{f_{l}^{(v)}}{\omega_{v}}\right]\label{eq:unitary_transform}\\
 & +\sum_{n=0}^{N-1}\sum_{v}|G,1_{ph}\rangle\langle G,1_{ph}|\left[(b_{v,n}^{\dagger}-b_{v,n})\frac{h^{(v)}}{\omega_{v}}\right]\nonumber \\
 & +\sum_{n=0}^{N-1}\sum_{v}|T_{n}\rangle\langle T_{n}|(b_{v,n}^{\dagger}-b_{v,n})\frac{f_{T}^{(v)}}{\omega_{v}}\nonumber \\
\equiv & \sum_{n=0}^{N-1}\left[|n\rangle\langle n|\hat{D}_{S}^{(n)}+|G,1_{ph}\rangle\langle G,1_{ph}|\hat{D}_{ph}+|T_{n}\rangle\langle T_{n}|\hat{D}_{T}^{(n)}\right]\nonumber 
\end{align}
is partially based on previous works \cite{Wu2016,Zeb2018}. According
to the ansatz in Eq. (\ref{eq:unitary_transform}), the electronic
and photonic degrees of freedom are dressed with vibrational bath
excitations to an extent quantified by the set of parameters $\{f_{l}^{(v)},f_{T}^{(v)},h^{(v)}\}$
which are variationally determined, as will be explained below. The
summation over $l$ in Eq. (\ref{eq:unitary_transform}) assumes that
the vibrational deformation can be extended over all sites in the
singlet excited manifold (large polaron limit) as a result of the
interaction of all optically bright singlet excitons to the same photon
mode. Since the electronic triplet states do not directly interact
with the photonic mode, the vibrational deformation of the triplet
excited manifold is expected to be more localized (small polaron limit).
In the polaron frame we have

\begin{align}
e^{P}He^{-P} & =\tilde{H}_{0}+\tilde{H}_{I}+H_{B},\label{eq:partition}
\end{align}
where

\begin{align}
\tilde{H}_{0}= & \sum_{n}\tilde{\epsilon}_{S}|n\rangle\langle n|+\sum_{n}\tilde{\epsilon}_{T}|T_{n}\rangle\langle T_{n}|+\tilde{\omega}_{ph}|G,1_{ph}\rangle\langle G,1_{ph}|\nonumber \\
 & +g\eta_{S-ph}\sum_{n}\left(|n\rangle\langle G,1_{ph}|+\text{h.c.}\right)\nonumber \\
 & +V_{ST}\eta_{ST}\sum_{n}\left(|n\rangle\langle T_{n}|+\text{h.c.}\right).\nonumber \\
= & \tilde{H}_{0,k=0}+\sum_{k\neq0}\tilde{H}_{0,k}\label{eq:zeroth}
\end{align}
and

\begin{subequations}\label{eq:H0_k}

\begin{align}
\tilde{H}_{0,k=0}= & \tilde{\epsilon}_{S}|k=0\rangle\langle k=0|+\tilde{\epsilon}_{T}|T_{k=0}\rangle\langle T_{k=0}|\nonumber \\
 & +\tilde{\omega}_{ph}|G,1_{ph}\rangle\langle G,1_{ph}|\nonumber \\
 & +\frac{\Omega}{2}\eta_{S-ph}\left(|k=0\rangle\langle G,1_{ph}|+\text{h.c.}\right)\nonumber \\
 & +V_{ST}\eta_{ST}\left(|k=0\rangle\langle T_{k=0}|+\text{h.c.}\right),\label{eq:H_0_k_0}\\
\tilde{H}_{0,k\neq0}= & \tilde{\epsilon}_{S}|k\rangle\langle k|+\tilde{\epsilon}_{T}|T_{k}\rangle\langle T_{k}|+V_{ST}\eta_{ST}\left(|k\rangle\langle T_{k}|+\text{h.c.}\right).\label{eq:H_0_k_neq_0}
\end{align}

\end{subequations}\noindent  In Eqs. (\ref{eq:zeroth}-\ref{eq:H0_k})
we introduced a delocalized Fourier basis for the singlet $|k\rangle=\frac{1}{\sqrt{N}}\sum_{n}e^{ikn}|n\rangle$
and triplet excitons $|T_{k}\rangle=\frac{1}{\sqrt{N}}\sum_{n}e^{ikn}|T_{n}\rangle$,
with $k=\frac{2\pi}{N}$, $m=0,1,2,\dots,N-1$. In this basis, the
state $|k=0\rangle$ couples to the photonic mode with a superradiantly
enhanced strength given by $\sqrt{N}g=\Omega/2$. The renormalized
on-site energies in (\ref{eq:zeroth}) are given by

\begin{subequations}\label{renorm}

\begin{align}
\tilde{\epsilon}_{S} & =\epsilon_{S}+\sum_{v}\left(\sum_{l}\frac{f_{l}^{(v)2}}{\omega_{v}}-2g_{S}^{(v)}\frac{f_{0}^{(v)}}{\omega_{v}}\right)\label{eq:S_renorm}\\
\tilde{\epsilon}_{T} & =\epsilon_{T}+\sum_{v}\left(\frac{f_{T}^{(v)2}}{\omega_{v}}-2g_{T}^{(v)}\frac{f_{0}^{(v)}}{\omega_{v}}\right)\label{eq:T_renorm}\\
\tilde{\omega}_{ph} & =\omega_{ph}+\sum_{v}N\frac{h^{(v)2}}{\omega_{v}}\label{eq:Ph_renorm}
\end{align}
\end{subequations}\noindent and the renormalization constant $\eta_{ST}$
($\eta_{S-ph}$) is the thermal average of the relative displacement
operator between the singlet and triplet (photonic) harmonic potential
energy surfaces, \begin{subequations}\label{coup_renorm}

\begin{align*}
\eta_{S-ph} & =\langle e^{\hat{D}_{S}^{(n)}}e^{-\hat{D}_{ph}}\rangle\\
 & =\exp\left[-\frac{1}{2}\sum_{l,v}\left(\frac{(f_{l}^{(v)}-h^{(v)})^{2}}{\omega_{v}^{2}}\right)\coth(\beta\omega_{v}/2)\right],\\
\eta_{ST} & =\langle e^{\hat{D}_{S}^{(n)}}e^{-\hat{D}_{T}^{(n)}}\rangle\\
 & =\exp\bigg\{-\frac{1}{2}\sum_{v}\bigg[\frac{(f_{0}^{(v)}-f_{T}^{(v)})^{2}}{\omega_{v}^{2}}\\
 & +\sum_{l\neq0}\frac{|f_{l}^{(v)}|^{2}}{\omega_{v}^{2}}\bigg]\coth(\beta\omega_{v}/2)\bigg\},
\end{align*}

\end{subequations}\noindent where $\langle A\rangle=\frac{\text{Tr}_{B}\{Ae^{-\beta H_{B}}\}}{Z_{H_{B}}}$,
$\text{Tr}_{B}\{\cdot\}$ denoting a trace over the vibrational degrees
of freedom, $\beta$ being the inverse temperature, and $Z_{H_{B}}=\text{Tr}_{B}\{e^{-\beta H_{B}}\}$
being the vibrational partition function. Notice that by taking $f_{l}^{(v)}=g_{S}^{(v)}\delta_{l,0}$
in Eq. (\ref{eq:S_renorm}), we recover the full-polaron transformation
for the singlet excitation, and $R_{S}=-\lambda_{S}=-\sum_{v}(g_{S}^{(v)})^{2}/\omega_{v}$,
where $\lambda_{S}$ is the reorganization energy of the singlet excited
state. In Eqs. (\ref{eq:zeroth})--(\ref{eq:H0_k}), we also introduced
a partition of the Hamiltonian into the molecular contribution that
is totally symmetric under site permutations $\tilde{H}_{0,k=0}$
and the non-totally-symmetric one $\sum_{k\neq0}\tilde{H}_{0,k}$.
Since $[\tilde{H}_{0,k=0},\sum_{k\neq0}\tilde{H}_{0,k}]=0,$ the eigenstates
of $\tilde{H}_{0}$ can be found by independent diagonalization of
each contribution: 
\begin{align}
\tilde{H}_{0,k=0} & =\tilde{\epsilon}_{\text{UP}}|\text{UP}\rangle\langle\text{UP}|+\tilde{\epsilon}_{\text{MP}}|\text{MP}\rangle\langle\text{MP}|+\tilde{\epsilon}_{\text{LP}}|\text{LP}\rangle\langle\text{LP}|,\label{eq:H_0_k_0-1}
\end{align}
where we define the upper (UP), middle (MP) and lower (LP) polariton
states $|j\rangle=c_{S}^{j}|k=0\rangle+c_{T}^{j}|T_{k=0}\rangle+c_{ph}^{j}|G,1_{ph}\rangle$,
$j=\text{UP},\thinspace\text{MP},\thinspace\text{LP}$. On the other
hand, 
\begin{equation}
\sum_{k\neq0}\tilde{H}_{0,k}=\sum_{k}\Big[\tilde{\epsilon}_{+}|k^{(+)}\rangle\langle k^{(+)}|+\tilde{\epsilon}_{-}|k^{(-)}\rangle\langle k^{(-)}|\Big]\label{Dark_states}
\end{equation}
describes mixed singlet-triplet eigenstates $|k^{(\pm)}\rangle=c_{S}^{\pm}|k\rangle+c_{T}^{\pm}|T_{k}\rangle$
with no photonic component, whose eigenenergies are given by $\tilde{\epsilon}_{\pm}=\frac{\tilde{\epsilon}_{S}+\tilde{\epsilon}_{T}}{2}\pm\frac{1}{2}\sqrt{(\tilde{\epsilon}_{S}-\tilde{\epsilon}_{T})^{2}+4V_{ST}^{2}\eta_{ST}^{2}}$.
These states are commonly known as dark or exciton reservoirs \cite{Agranovich2003,Litinskaia2004}.
For clarity, we refer to the highest (lowest) energy eigenstates in
Eq. (\ref{Dark_states}) as the upper (lower) dark states. Notice
that since we are ignoring inter-site couplings, we neglect the dispersive
character of the exciton in the calculation of the eigenenergies (\ref{Dark_states}),
a valid assumption in view of the flat exciton dispersion relation
compared to the photonic one in the $k$ range of interest. The residual
interaction Hamiltonian in Eq. (\ref{eq:partition}) is 
\begin{align}
\tilde{H}_{I} & =\tilde{V}_{S}+\tilde{V}_{T}+\tilde{V}_{ph}+\tilde{V}_{S-T}+\tilde{V}_{S-ph},\label{res_int}
\end{align}
which has been written as a sum of different contributions defined
as follows, 

\begin{subequations}\label{residual}

\begin{align}
\tilde{V}_{S}= & \sum_{v}\sum_{n}|n\rangle\langle n|\bigg[(g_{S}^{(v)}-f_{0}^{(v)})(b_{v,n}^{\dagger}+b_{v,n})\nonumber \\
 & -\sum_{l\neq0}f_{l}^{(v)}(b_{v,n+l}^{\dagger}+b_{v,n+l})\bigg],\label{eq:res_S}\\
\tilde{V}_{T}= & \sum_{v}\sum_{n}|T_{n}\rangle\langle T_{n}|\left[(g_{T}^{(v)}-f_{T}^{(v)})(b_{v,n}^{\dagger}+b_{v,n})\right],\label{eq:res_T}\\
\tilde{V}_{ph}= & -\sum_{v}h^{(v)}|G,1_{ph}\rangle\langle G,1_{ph}|\sum_{n}\left(b_{v,n}^{\dagger}+b_{v,n}\right),\label{eq:res_ph}\\
\tilde{V}_{S-T}= & \sum_{n}V_{ST}\left[e^{\hat{D}_{S}^{(n)}-\hat{D}_{T}^{(n)}}-\eta_{TS}\right]|n\rangle\langle T_{n}|+\text{h.c.},\label{eq:res_VST}\\
\tilde{V}_{S-ph}= & \sum_{n}g\left[e^{\hat{D}_{S}^{(n)}-\hat{D}_{ph}^{(n)}}-\eta_{S-ph}\right]|n\rangle\langle G,1_{ph}|+\text{h.c.}.\label{eq:res_Sph}
\end{align}

\end{subequations}\noindent The calculation of the parameters $\{f_{l}^{(v)},f_{T}^{(v)},h^{(v)}\}$
is carried out by taking advantage of the Feynman-Bogoliubov inequality
$F\le F'_{0}+\langle\tilde{H}_{I}\rangle_{\tilde{H}_{0}'}$ \cite{Silbey1984,Harris1985},
where $F$ ($F'_{0}$) is the free energy of the system governed by
$H$ ($\tilde{H}_{0}'=\tilde{H}_{0}+H_{B}$) and $\langle A\rangle_{\tilde{H}_{0}'}=\text{Tr}\{Ae^{-\beta\tilde{H}_{0}'}\}[\text{Tr}\{e^{-\beta\tilde{H}_{0}'}\}]^{-1}$
where $\text{Tr}\{\cdot\}$ denotes the trace over the eigenstates
of $\tilde{H}_{0}'$ . Notice that by construction $\langle\tilde{H}_{I}\rangle_{\tilde{H}'_{0}}=0$,
and the leading correction to the exact equilibrium reduced density
matrix is $O(\tilde{H}_{I}^{2})$ \cite{Lee2012}, which justifies
the (second order) perturbative treatment in $\tilde{H}_{I}$. It
follows that the parameters $\{f_{l}^{(v)},f_{T}^{(v)},h^{(v)}\}$
can be found by minimizing $F'_{0}=-\beta^{-1}\text{Tr}\{e^{-\beta\tilde{H}_{0}}\}$,
which amounts to solving $\frac{\partial F'_{0}}{\partial f_{i}^{(v)}}=\frac{\partial F'_{0}}{\partial h^{(v)}}=0$.
In our calculations, we consider a continuum vibrational bath limit,
which is described in terms of the spectral densities $J_{i}(\omega)=\sum_{v}|g_{v}^{i}|^{2}\delta(\omega-\omega_{v})=a_{i}\frac{\omega^{3}}{\omega_{c}^{2}}e^{-\omega/\omega_{c}}$,
$i=S,T$, where $\omega_{c}$ is a cutoff frequency, and $a_{i}$
is the dimensionless parameter that encodes the strength of coupling
between the excited electronic state with character $i$ to the vibrational
bath. The computational details of the solution for $\{f_{l}^{(v)},f_{T}^{(v)},h^{(v)}\}$
are summarized in the Appendix A. In the discussion that follows,
we focus on the calculation of the population dynamics of the chemically
relevant species as well as of the electroluminescence efficiencies.

\section{Dynamics in the polaron frame and definition of triplet electroluminescence
efficiency\label{sec:Dynamics-in-the}}

The open-quantum system dynamics associated with the Hamiltonian in
Eq. (\ref{eq:Hamiltonian}) can be described in terms of the time
evolution of the reduced density matrix (RDM) $\rho(t)=\text{Tr}_{B}\{\rho_{tot}(t)\}$,
($\rho_{tot}(t)$ is the total density matrix), governed by the Liouville
equation in the Schrödinger picture 
\begin{align*}
\frac{d\rho(t)}{dt} & =\mathcal{L}\rho(t)-\{\Gamma_{nrad}+\Gamma_{ph},\rho(t)\},
\end{align*}
where $\mathcal{L}\rho(t)=-i\text{Tr}_{B}\{[H,\rho_{tot}(t)]\}$,
and $\{\rho(t),\Gamma_{i}\}=\rho(t)\Gamma_{i}+\Gamma_{i}\rho(t)$;
$\Gamma_{rad}$, $\Gamma_{nrad}$, $\Gamma_{ph}$ phenomenologically
account for the dissipative processes associated with the radiative
and nonradiative decay of excitons \cite{Rebentrost2009_JPCB}, and
photonic leakage, respectively. We define the triplet electroluminescence
efficiency 
\begin{equation}
\epsilon=2\int_{0}^{\infty}\text{Tr}\left\{ \Gamma_{ph}\rho(\tau)\right\} d\tau,\label{eq:tot_eff}
\end{equation}
where the trace is taken over all the degrees of freedom, namely,
the electronic, vibrational, and photonic. Eq. (\ref{eq:tot_eff})
is analogous to the integrated probability used to define the efficiency
of energy trapping in chromophoric complexes \cite{Rebentrost2009_JPCB},
but in the present context, acquires the meaning of the efficiency
of emission of a photon. Here, we define $\Gamma_{nrad}=\frac{k_{nrad}^{S}}{2}\sum_{n}|n\rangle\langle n|+\frac{k_{nrad}^{T}}{2}\sum_{n}|T_{n}\rangle\langle T_{n}|$
and $\Gamma_{ph}=\frac{\kappa}{2}|G,1_{ph}\rangle\langle G,1_{ph}|$,
$k_{nrad}^{i}$ being the nonradiative decay rate of electronic state
$i=S,T$, and $\kappa$ being the cavity photon leakage rate. Since
\begin{align}
\text{Tr}\{\Gamma_{ph}\rho_{tot}(t)\}= & \text{Tr}\{\Gamma_{ph}e^{-P}\tilde{\rho}_{tot}(t)e^{P}\}\nonumber \\
= & \text{Tr}\{\Gamma_{ph}\tilde{\rho}_{tot}(t)\},\label{eq:trace}
\end{align}
$\epsilon$ is an invariant quantity under the polaron transformation.
This last condition permits the computation of Eq. (\ref{eq:tot_eff})
in the polaron frame, where the time evolution of the polaron transformed
RDM $\tilde{\rho}(t)=\text{Tr}_{B}\{\tilde{\rho}_{tot}\}$ can be
described in terms of second-order perturbation theory on $\tilde{H}_{I}$
within the secular Born-Markov approximation. This procedure guarantees
that the long-time evolution of the polaritonic system properly thermalizes
into the reduced equilibrium state due to $\tilde{H}_{0}$ which,
due to the optimization of the variational parameters according to
the Feynman-Bogoliubov bound (see Sec. \ref{sec:Theory}), provides
a good description of the equilibrium state of the full system involving
electronic states, vibrations, and photon. 

For the calculation of $\epsilon$, we assume that the initial RDM
corresponds to an incoherent mixture of localized excitations in the
triplet electronic manifold, \emph{i.e., $\sum_{n}\langle T_{n}|\tilde{\rho}(0)|T_{n}\rangle=1$,
$\sum_{n}\langle T_{n}|\tilde{\rho}(0)|n\rangle=\sum_{n}\langle T_{n}|\tilde{\rho}(0)|G,1_{ph}\rangle=\sum_{n}\langle n|\tilde{\rho}(0)|n\rangle=\langle G,1_{ph}|\tilde{\rho}(0)|G,1_{ph}\rangle=0$.
}Under these assumptions, the initial RDM in the polaron frame can
be written as 
\begin{equation}
\vec{\tilde{\rho}}(0)\approx\begin{bmatrix}\langle\text{UP}|\tilde{\rho}(0)|\text{UP}\rangle\\
\langle\text{MP}|\tilde{\rho}(0)|\text{MP}\rangle\\
\langle\text{LP}|\tilde{\rho}(0)|\text{LP}\rangle\\
\sum_{k\neq0}\langle k^{+}|\tilde{\rho}(0)|k^{+}\rangle\\
\sum_{k\neq0}\langle k^{-}|\tilde{\rho}(0)|k^{-}\rangle
\end{bmatrix}\approx\begin{bmatrix}0\\
0\\
0\\
|c_{T}^{+}|^{2}\\
|c_{T}^{-}|^{2}
\end{bmatrix},\label{eq:initial_state}
\end{equation}
where the approximation is the result of considering a localized initial
state, which in the delocalized polaron picture translates into the
population distributed uniformly among the dark and polariton states.
Since the former feature a larger DOS than the latter, the population
is predominantly concentrated in the dark states. We can rewrite Eq.
(\ref{eq:tot_eff}) as 
\begin{equation}
\epsilon=\int_{0}^{\infty}\kappa P_{ph}\cdot\vec{\tilde{\rho}}(\tau)d\tau,\label{eq:eff_def}
\end{equation}
where the vector 
\begin{align*}
P_{ph} & =\begin{bmatrix}|c_{ph}^{UP}|^{2} & |c_{ph}^{MP}|^{2} & |c_{ph}^{LP}|^{2} & 0 & 0\end{bmatrix}
\end{align*}
accounts for the photonic character of the different chemical species
that take part in the population dynamics. The time evolution of $\vec{\tilde{\rho}}(\tau)$
is described in terms of a Pauli master equation in the secular Markovian
approximation, where 
\begin{align}
\frac{d\vec{\tilde{\rho}}(t)}{dt} & =\mathcal{M}\vec{\tilde{\rho}}(t).\label{eq:Pauli}
\end{align}
For simplicity in the calculations and interpretation, we consider
the evolution of the total population in the dark state manifolds
$\sum_{k\neq0}\langle k^{\pm}|\tilde{\rho}(t)|k^{\pm}\rangle$ rather
than dissected among the individual populations $\{\langle k^{\pm}|\tilde{\rho}(t)|k^{\pm}\rangle\}$.
The matrix $\mathcal{M}$ is given by 
\begin{equation}
\mathcal{M}_{ij}=\begin{cases}
k_{i\leftarrow j} & i\neq j\\
-\sum_{l\neq i}k_{l\leftarrow i}-\kappa_{phot}|\langle i|G,1_{ph}\rangle|^{2} & i=j\\
-k_{nrad}^{T}\langle i|\mathcal{I}_{T}|i\rangle-k_{nrad}^{S}\langle i|\mathcal{I}_{S}|i\rangle
\end{cases}\label{eq:rate_Matrix}
\end{equation}
where $k_{j\leftarrow i}$ is the rate of transfer from the state/manifold
$|i\rangle$ to $|j\rangle$, and $\mathcal{I}_{T}=\sum_{n}|T_{n}\rangle\langle T_{n}|$,
$\mathcal{I}_{S}=\sum_{n}|n\rangle\langle n|$. Details of the calculation
of the population transfer rates are included in the Appendix B. Using
Eq. (\ref{eq:Pauli}) together with Eq. (\ref{eq:eff_def}), we have
that 
\begin{align}
\epsilon & =-\kappa P_{ph}\mathcal{M}^{-1}\vec{\tilde{\rho}}(0).\label{eq:toteff}
\end{align}
To get further insight into the contributions of the different pathways
of population transfer to $\epsilon$, we consider a Green's function
partition approach \cite{Rebentrost2009_JPCB}, under which 
\begin{equation}
\mathcal{M}^{-1}=\mathcal{M}_{npol}^{-1}-\mathcal{M}_{npol}^{-1}\mathcal{M}_{pol}\mathcal{M}^{-1},\label{eq:part_pol}
\end{equation}
where we let $\mathcal{M}=\mathcal{M}_{pol}+\mathcal{M}_{npol}$,
$\mathcal{M}_{pol}$ being the rate matrix that accounts for the feed
and depletion of polariton states only, which amounts to considering
the rates $k_{i\leftarrow j}$ where either $i$ or $j\in\{|UP\rangle,|MP\rangle,|LP\rangle\}$
while setting the other entries of $\mathcal{M}$ to $0$. Using Eqs.
(\ref{eq:toteff}) and (\ref{eq:part_pol}), 
\[
\epsilon=\epsilon_{dir}+\epsilon_{indir}
\]
where $\epsilon_{dir}=\kappa P_{ph}\mathcal{M}_{npol}^{-1}\mathcal{M}_{pol}\mathcal{M}^{-1}\vec{\tilde{\rho}}(0)$,
$\epsilon_{indir}=-\kappa P_{ph}\mathcal{M}_{npol}^{-1}\vec{\tilde{\rho}}(0)$,
and $\epsilon$ can be written as a sum of a contribution due to polariton-participating
processes and another one where they do not play a role.

\section{Results and discussion\label{sec:Results-and-discussion}}

Based on the approach above, we are particularly interested in the
dependence of $\epsilon$ with Rabi energy $\Omega$ and detuning
between the cavity photon and the vertical singlet energy transition
$(\Delta=\omega_{ph}-\epsilon_{S})$, considering molecular emitters
which feature different regimes of interaction with the vibrational
bath degrees of freedom (DOFs). For that purpose, we introduce two
cases based on parameters chosen to represent a wide range of organic
molecular emitters. 

First, we consider (large) strengths of coupling to the vibrational
environment $a_{S}=10$, $a_{T}=15$, and a frequency cutoff $\omega_{c}=0.01$
eV. The corresponding reorganization energy of the singlet electronic
state is given by $\lambda_{S}=\int_{0}^{\infty}\frac{J_{S}(\omega)}{\omega}d\omega=0.2$
eV and the analogous quantity for the triplet is $\lambda_{T}=\int_{0}^{\infty}\frac{J_{T}(\omega)}{\omega}d\omega=0.3$
eV. The assumption of a single vibrational reservoir coupled to both
the singlet and triplet states allows for the calculation of the reorganization
energy of the singlet-triplet transition $\lambda_{ST}$ (see Appendix
A for details), which for this specific case amounts to 0.01 eV. For
simplicity, we do not consider high frequency vibrational bath modes
(those which feature frequencies $\omega\gg\beta^{-1}$) in our calculations.
Furthermore, we assume an energy gap between the equilibrium vibrational
configurations of the triplet and singlet $\Delta G=(\epsilon_{T}-\lambda_{T})-(\epsilon_{S}-\lambda_{S})=-0.1$
eV and a weak intersystem crossing coupling amplitude $V_{ST}=2\times10^{-5}$
eV. These parameters are typical for organic molecules that undergo
thermally activated delayed fluorescence (TADF) \cite{Chen2015,yang2017,Peng2017},
where the population in the dark triplet states transfers to the bright
singlets, with a R-ISC rate that depends on a thermal energy barrier
between the latter. We also consider a relatively slow nonradiative
decay of the triplet $k_{nrad}^{T}=2\times10^{-7}$ ps$^{-1}$ \cite{yang2017}.
Since the characteristic timescale of relaxation of the vibrational
environment $\omega_{c}^{-1}\ll\hbar/|V_{ST}|$, we expect a fast
relaxation of the latter into its equilibrium configuration before
R-ISC ensues, a scenario termed 'fast bath' in the literature \cite{Lee2012}.
This is the behaviour expected in the limit $\Omega\rightarrow0$
(small polaron formation before electronic transition) which was confirmed
in our numerical calculations.

The second case we considered is the opposite to the previous one,
in which the vibrational DOFs of the environment are sluggish in comparison
to the R-ISC transition time scale ($V_{ST}\gg\omega_{c}$), a scenario
identified as 'slow bath' \cite{Lee2012}, which corresponds to molecular
emitters with sizable singlet-triplet mixing via spin-orbit coupling
or hyperfine interaction. For this case, we also chose large strengths
of coupling to the vibrational environment $a_{S}=10$, $a_{T}=30$,
while setting a small cutoff frequency $\omega_{c}=5\times10^{-3}$
eV (which translate into $\lambda_{S}=0.01$ eV, $\lambda_{T}=0.03$
eV and $\lambda_{ST}=0.1$ eV). Furtheremore, we set $V_{ST}=0.05$
eV and $\Delta G=-0.01$ eV, parameters that are qualitatively aligned
with molecules that exhibit fast singlet-fission \cite{Yost2014},
which exhibit large electronic couplings, comparable to the energy
gaps between the singlet and the (two-body) triplet state. For this
case, if $k_{nrad}^{T}$ is small as above, the cavity-free triplet
electroluminescence efficiency $\epsilon\approx1$ and placing it
in the cavity is not productive; hence, we assume that the molecule
is such that it features a fast nonradiative triplet decay $k_{nrad}^{T}=2\times10^{-3}$
ps$^{-1}$ , such that we explore the possibility to outcompete it
by cavity-assisted processes. The temperature $T=300$ K is assumed
for both cases.

For reference, we computed electroluminescence efficiencies in the
cavity-free (bare molecules) case which are $\epsilon=0.027$ for
the fast bath case, and $\epsilon=0.062$ for the slow one. The latter
are calculated using Eq. (\ref{eq:tot_eff}), substituting $\Gamma_{ph}$
with $\Gamma_{rad}=\frac{k_{rad}^{S}}{2}\mathcal{I}_{S}$ , where
$k_{rad}^{S}=10^{-2}$ ps$^{-1}$, a typical fluorescence rate for
organic molecules. 

\begin{figure*}
\includegraphics[width=0.8\textwidth]{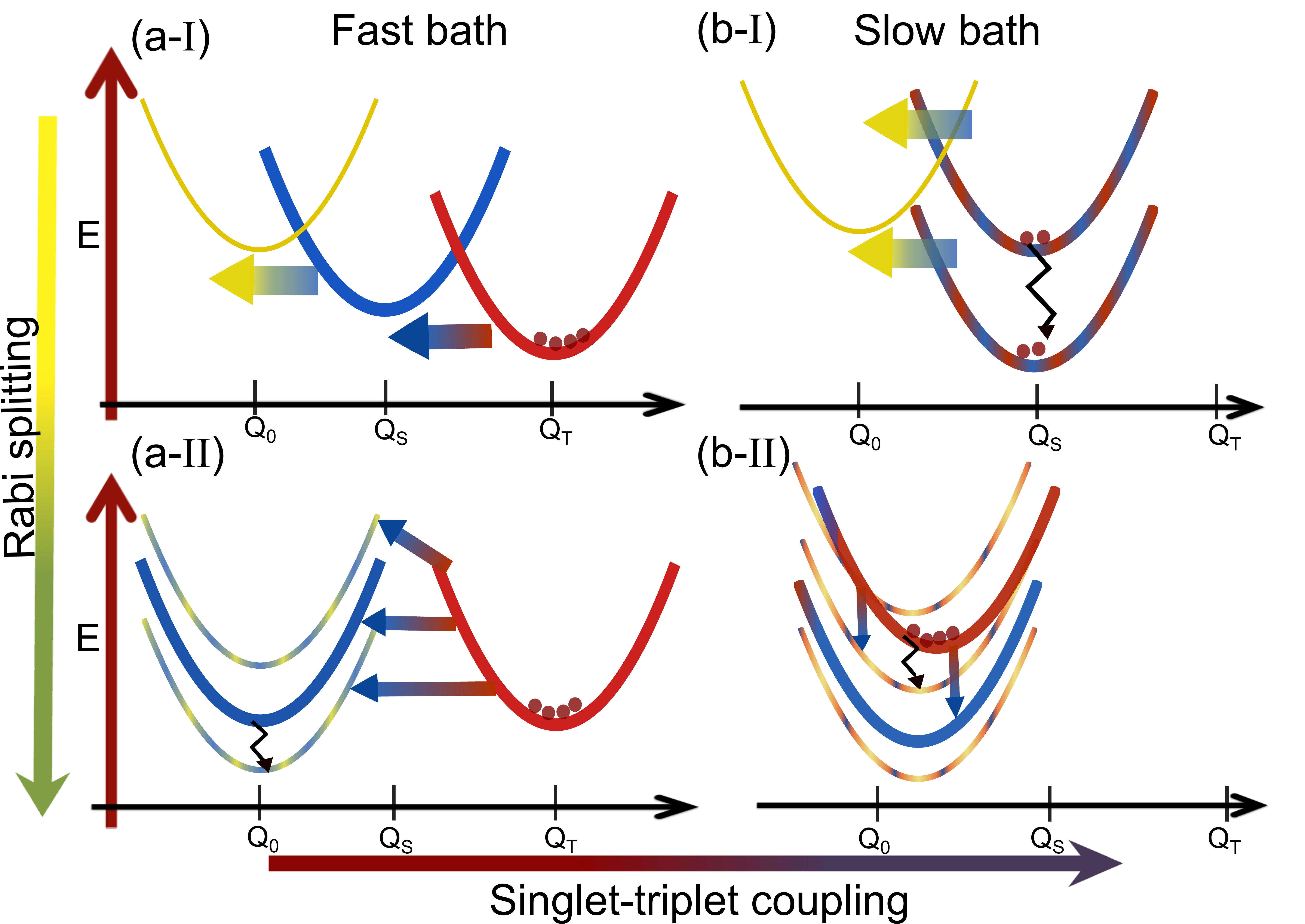}

\caption{Scheme of polaron frame population dynamics which emerge upon injection
of optically dark triplet excitons into molecules confined in an optical
microcavity. For simplicity, we depict only one vibrational mode and
represent its harmonic potential energy E as a function of its coordinate
$Q$, in the photonic state (molecular ground state with one photon,
yellow thin curve), and corresponding harmonic potential manifolds
for singlet (blue thick curve) and triplet (red thick curve) excited
states due to $N$ molecules. We consider molecules which feature
(a) fast and (b) slow vibrational environment (see main text), compared
to the R-ISC timescale defined by the singlet-triplet coupling $\hbar/V_{ST}$.
For each case, we find qualitatively different mechanisms of population
transfer among states for (I) weak and (II) strong light-matter coupling.
The most significant mechanisms are depicted: multiphonon or Marcus-like
processes with straight arrows, and single-phonon or Redfield-like
processes with jagged ones.\label{fig:Scheme_dynamics}}
\end{figure*}

\begin{figure*}
\centering{}\includegraphics[width=1\textwidth]{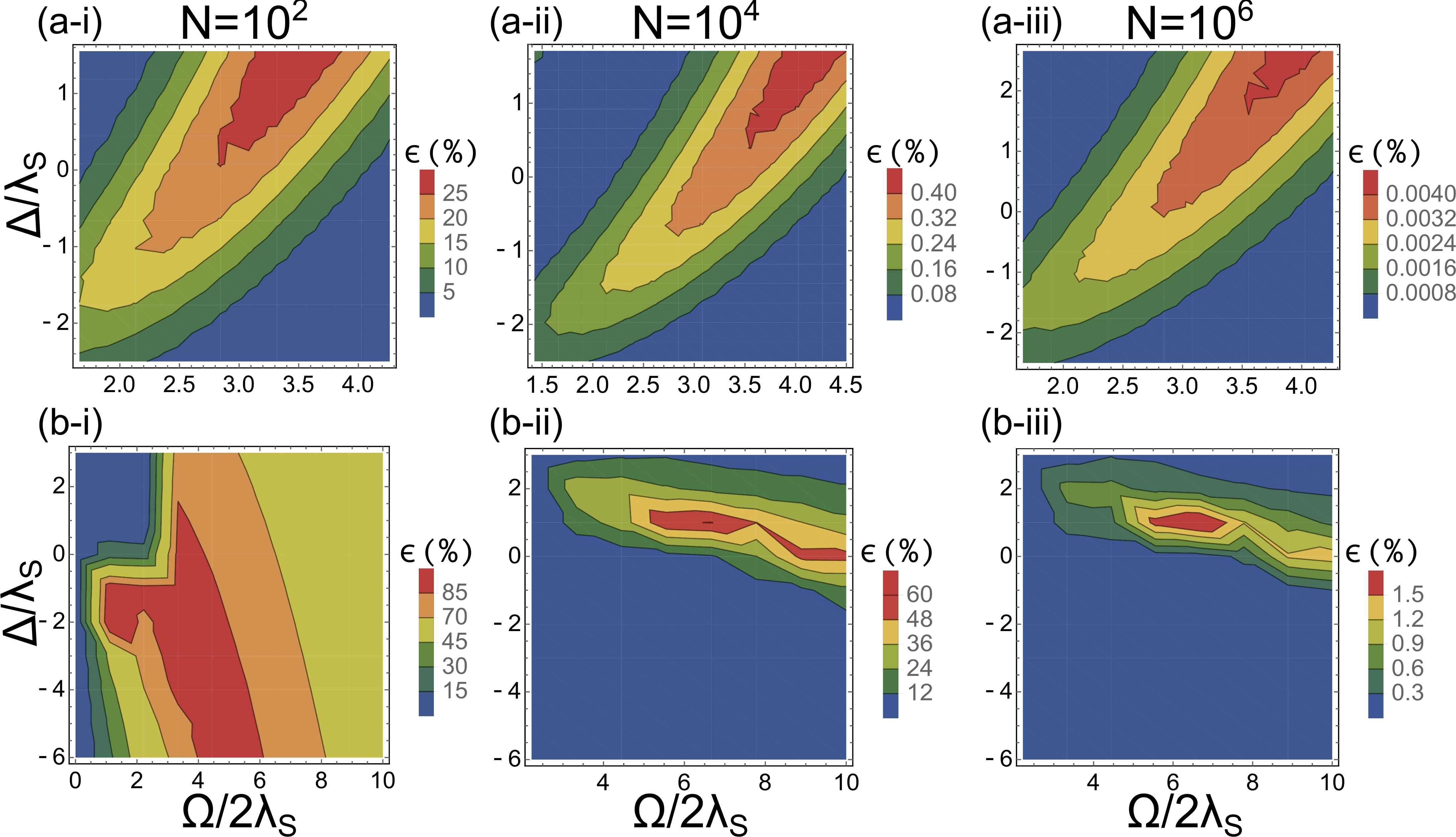}
\caption{Triplet electroluminescence efficiency $\epsilon$ as a function of
cavity detuning $\Delta=\omega_{ph}-\epsilon_{S}$ and Rabi splitting
$\Omega$ for (a) the fast bath case, where vibrational equilibration
occurs before a slow R-ISC and b) the slow bath case, where the opposite
is true (where the emitters feature strong singlet-triplet mixing,
see main text). The yields outside of the cavity are (a) $\epsilon=0.027$
for the fast bath and (b) $\epsilon=0.062$ for the slow one. The
effective number of molecules per photon mode is: (i) $N=10^{2}$,
(ii) $N=10^{4}$ and (iii) $N=10^{6}$. \label{fig:2Dplots}}
\end{figure*}

Upon coupling to a cavity photon mode, the emergent dynamics are changed
as a result of the modification of the bare molecular energy landscape
and the reorganization energies between the polariton states. In the
polaron frame, the former contribution is encoded in the energy spectrum
introduced in Eqs. (\ref{eq:H_0_k_0-1}-\ref{Dark_states}), whereas
the latter is accounted for in the reorganization energies $\lambda_{i-j}$,
$i\neq j$, $i,j\in UP,MP,LP,\pm$, where the subindices $\pm$ denote
${|k^{\pm}\rangle}$ states (in our model, the reorganization energies
$\lambda_{i-(\pm)}$ are independent of the momentum of the dark states).
The polaron frame picture of the dynamics in both the slow and the
fast bath case is schematically illustrated in Fig. \ref{fig:Scheme_dynamics}
and the variation of the electroluminescence efficiency with $\Delta$
and $\Omega$ for the two different molecular emitter cases is shown
in Fig. \ref{fig:2Dplots}.

For the discussion in this section, we limit ourselves to the $N=10^{2}$
case, postponing the important analysis of larger $N$ values for
the next paragraphs. 

\begin{figure*}
\includegraphics[width=0.8\textwidth]{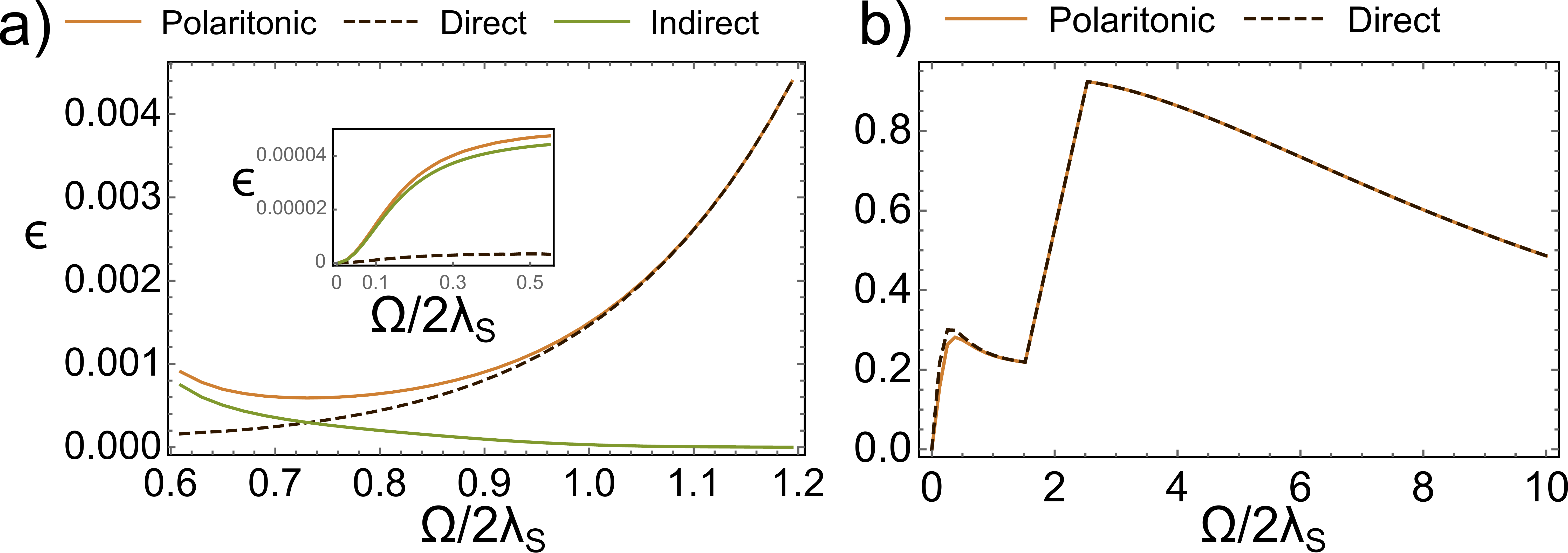}
\caption{Partition of triplet electroluminescence efficiencies as a function
of Rabi splitting $\Omega$ at light-matter resonance $\omega_{ph}=\epsilon_{S}$
(for $\Delta=0$) for a) a fast bath and b) a slow bath molecule;
we show the case for $N=100$. The triplet electroluminescence efficiency
$\epsilon$ (labeled as polaritonic) can be partitioned into a fraction
that accounts for the direct and indirect population transfer from
dark triplets into polaritons, where indirect means that population
passes through other dark states before it reaches the polaritons.
In b), we do not explicitly show the indirect contribution since it
is negligible for featured Rabi splitting ranges. Inset in (a) zooms
into smaller Rabi splitting ranges. For comparison, the yields outside
of the cavity are (a) $\epsilon=0.027$ for the fast bath and (b)
$\epsilon=0.062$ for the slow one. \label{fig:Partitions}}
\end{figure*}

\emph{Fast bath scenario.-} For the fast bath molecule, we find that
for weak light-matter couplings (Rabi splittings $\Omega\ll\lambda_{S}$),
$\epsilon(\Delta,\Omega)$ is below the cavity-free scenario (for
the resonant case $\Delta=0$, see inset in Fig. \ref{fig:Partitions}a).
In the polaron frame, this observation can be explained in terms of
thermally-activated transfer of population from the triplets to the
photonic potential energy surface (PES) via the singlets (see Fig.
\ref{fig:Scheme_dynamics}a-I). Since this process involves passage
through two thermal energy barriers, the time needed to reach the
photonic PES is longer that the triplet lifetime and therefore $\epsilon\ll1$.
On the other hand, when $\Omega/2\lambda_{S}\approx2/3$, our calculations
predict an abrupt increase in $\epsilon$ (compare inset and main
plot in Fig \ref{fig:Partitions}a). This is a manifestation of polaron
decoupling \cite{Herrera2017,Zeb2018} that is expected for sufficiently
large Rabi splittings \cite{Herrera2016,Zeb2018}, under which the
polariton PESs are essentially undisplaced with respect to the electronic
ground PES. The intuition behind this decoupling is that the exchange
of energy between photon and singlet excitons is much faster than
that between vibrations and singlet excitons. Importantly, while this
regime was previously predicted for $\Omega\gg\lambda_{S}$ \cite{Herrera2018}
for a single high-frequency mode, we hereby numerically demonstrate
that this stringent condition can be relaxed to smaller Rabi splittings
in the case of a continuum of low-frequency modes. In this polaron
decoupling regime, our calculations predict that $\lambda_{(-)-UP}=\lambda_{(-)-MP}=\lambda_{(-)-LP}=\lambda_{(+)-(-)}>\lambda_{ST}$
{[}where the equality among the previous quantities is a result of
the ansatz in Eq. (\ref{eq:unitary_transform}), which for $\Omega/2\lambda_{S}>2/3$,
corresponds to $f_{l}^{(v)}=f^{(v)}=O(\frac{g_{S}^{(v)}}{N})$, for
all $l$. In other words, the singlet exciton PESs are negligibly
displaced with respect to the ground state PES when $N$ is large{]}.
As a consequence, the R-ISC energy barrier can either decrease or
increase with respect to the bare molecules by tuning the Rabi splitting.
In fact, we notice that as $\Omega$ increases (see Figs. \ref{fig:2Dplots}a-i
and \ref{fig:Scheme_dynamics}a-II), $\epsilon$ increases and then
decreases with respect to the cavity-free value, given that the smallest
R-ISC to polariton energy barrier decreases and then increases as
this rate goes from normal to inverted Marcus regimes. Notice that
Fig. \ref{fig:Partitions}a only shows an increasing behavior in $\epsilon$
because the plotted range of Rabi splittings is small.

\emph{Slow bath scenario.-} For the slow bath case, we observe a significantly
different $\epsilon(\Delta,\Omega)$. In Fig. {[}\ref{fig:2Dplots}b-i{]},
we show that there is no need to invoke Rabi splittings beyond the
reorganization energy to observe an enhancement in electroluminescence
efficiency.

Similarly to the previous fast bath scenario, our calculations reveal
two qualitatively different pictures of the population dynamics in
the polaron frame depending on the magnitude of $\Omega$. For sufficiently
weak light-matter coupling, the initial state is comprised of approximately
equally populated lower and upper dark state manifolds since $|c_{S}^{+}|^{2}\approx|c_{S}^{-}|^{2}$
{[}see Eq. (\ref{eq:initial_state}) and Fig. \ref{fig:Scheme_dynamics}b-I{]},
a consequence of the prevalence of the electronic coupling $V_{ST}$
over the weak coupling of the R-ISC transition to the phonon environment.
In this regime, $\epsilon$ can be explained in terms of a Marcus
picture where the photonic PES receives population from the upper
and lower dark-state manifolds with a rate proportional to a diabatic
coupling of order $|g|^{2}$, corresponding to single-molecule light-matter
coupling {[}see Eq. (\ref{eq:res_Sph}) and Fig. \ref{fig:Scheme_dynamics}b-I{]}.
High and small R-ISC energy barriers can be obtained by scanning across
$\Delta$, thus decreasing and increasing $\epsilon$, respectively;
for the computed range of $\Delta$ values, the dominant R-ISC pathway
is the one starting from the upper dark states (they are closer in
energy to the photonic PES, see Fig. \ref{fig:Scheme_dynamics}b-I).
In fact, for sufficiently small increase in $\Omega$, the triplet
electroluminescence efficiency $\epsilon$ also rises because $|g|^{2}$
increases; see Fig. \ref{fig:Partitions}b, which shows a cut of $\epsilon$
at light-matter resonance $\Delta=0$. In this plot, this mechanism
is operative up to a first maximum in $\epsilon$ at $\Omega\approx\lambda$,
upon which light-matter coupling changes from weak to moderate and
$\epsilon$ decreases. This effect can be understood as follows: as
$\Omega$ increases, it can compete with $V_{ST}$, thus reducing
the magnitude of the dressed $\tilde{V}_{ST}$, and concomitantly
reducing the mixing between singlets and triplets. The result is that
the lower and upper dark states become more triplet and singlet-like,
respectively. Therefore, the initial triplet population has more probability
of residing in the lower dark states, but these feature a much larger
energy barrier to the photonic mode, thus leading to slower R-ISC
and lower $\epsilon$. 

Keeping our discussion around Fig. \ref{fig:Partitions}b, our model
predicts a sharp change in the dynamics under strong light-matter
coupling starting at $\Omega\approx4\lambda_{S}$, with $\epsilon$
increasing with $\Omega$, featuring a second maximum in $\epsilon$
at $\Omega\approx6\lambda_{S}$, and then decreasing for larger $\Omega$.
An intriguing feature of this regime is that, for the chosen parameters,
the upper and lower dark states have largely triplet and singlet characters,
respectively; that is, the configuration of minimal energy of the
vibrational DOFs coupled to the singlet and triplet states in the
polaron frame is such that their energy ordering is inverted with
respect to that in the original frame (Fig. \ref{fig:Scheme_dynamics}b-II)
where $\epsilon_{S}-\lambda_{S}>\epsilon_{T}-\lambda_{T}$. This effect
is due to the slow relaxation of the bath that renders the vibrational
dressing of the states at nonequilibrium nuclear configurations close
to those accessed by vertical transitions from the ground state. Under
these circumstances, population is primarily initialized in the upper
dark states and transfers to the MP (Fig. \ref{fig:Scheme_dynamics}b-II)
which is closer in energy than the other polaritons or dark states.
Notice that in contrast to the weak light-matter coupling case, MP
has significant singlet, triplet, and photonic characters and the
population transfer mechanism can be essentially dissected into sizable
contributions from multiphonon (a Marcus-like contribution due to
displacement between the upper dark states and MP harmonic potentials)
and one-phonon (Redfield-like) processes. In our example, a steady
boost in $\epsilon$ is given by the Redfield processes which dominate
over the Marcus-like ones, given the large activation energy barriers
to be surmounted in the latter. These Redfield processes are primarily
due to the triplet character in both upper dark states and MP. After
the second maximum in $\epsilon$, further increase of $\Omega$ leads
to a phonon blockade and a concomitant decrease in $\epsilon$: the
MP energy keeps lowering but no one-phonon process is available to
mediate the transfer from the upper dark states (the vibrational DOS
decreases exponentially as a function of frequency in our chosen spectral
density).

\begin{figure*}
\includegraphics[width=0.8\textwidth]{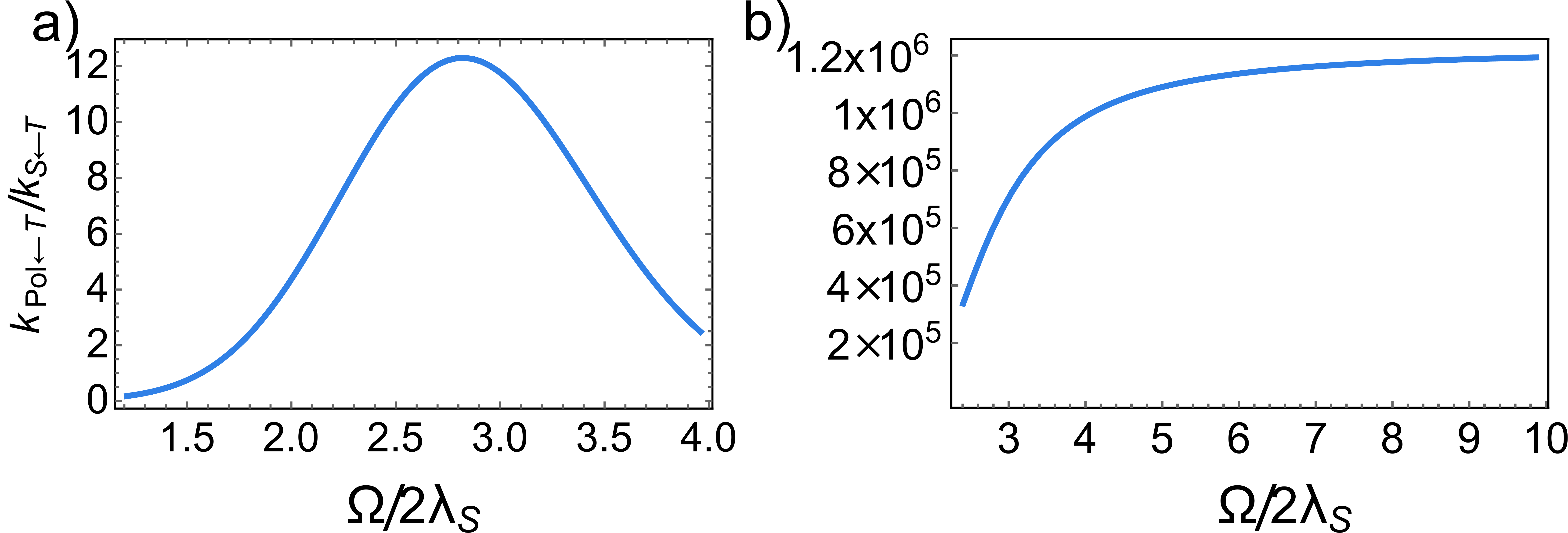}

\caption{Relative R-ISC rates from triplet manifold to polariton states (at
light-matter resonance $\omega_{ph}=\epsilon_{s}$) in the polaron
frame with respect to the cavity-free case, for (a) the fast bath
scenario and (b) the slow bath case, as a function of Rabi splitting,
considering $N=10^{2}$. The profile observed in (a) (where the bare
R-ISC rate is $k_{S\leftarrow T}=5.4\times10^{-9}$ ps$^{-1}$), can
be explained in terms of a multiphonon Marcus theory process from
triplets to singlets (see Fig. \ref{fig:Scheme_dynamics}a-II), where
the lowest activation barrier between triplets and polaritons decreases
(normal regime) and then increases (inverted regime). In (b) (where
the bare R-ISC rate is $k_{S\leftarrow T}=4.6\times10^{-6}$ ps$^{-1}$),
the trend is different since the dominant R-ISC channel is a one-phonon
process from triplets to the MP state (see Fig. \ref{fig:Scheme_dynamics}b-II)
and significantly depends on the triplet character of the latter.
The asymptotic behavior of the rate reflects the increase in triplet
character of the (MP) polariton state with Rabi splitting and the
suppression of the energy gap between the latter and the triplet states.
However, in this asymptotic limit, the photonic component of the MP
vanishes and this channel does not longer contribute to the cavity
electroluminescence efficiency. Finally, since the $k_{Pol\leftarrow T}$
rates shown here scale as $1/N$, the rates expected at different
$N$ can be calculated by scaling the values displayed accordingly.
\label{fig:rates_bothcases}}
\end{figure*}

\section{The large N issue\label{sec:largeN}}

So far, we have addressed the $N=100$ case in detail. In fact, strong
light-matter coupling with such small number of molecular emitters
has been reported using plasmonic nanoparticle environments \cite{halas,Bisht2018}.
However, most microcavity systems have much larger $N$ values. We
thus proceed to comment on the $\epsilon$ behavior for increasing
values of $N$, while keeping Rabi splittings $\Omega$ fixed (and
hence energy barriers and energy differences between states, see Fig.
\ref{fig:2Dplots}). For weak light-matter coupling, both fast (a)
and slow (b) baths yield insensitive changes of $\epsilon$ to increase
of $N$. This is due to the fact that changing $N$ does not alter
the energy barriers to be surmounted in the R-ISC process.

For strong light-matter coupling, the overall electroluminescence
efficiency is damped as a result of the ratio of triplet to polariton
states increasing. For the fast bath case (a), since both dark and
polariton states feature the same reorganization energies in the polaron
frame, $\epsilon$ is determined by the competition between the nonradiative
triplet decay and transfer to the polariton state closer in energy,
that is because the latter channel features a lower activation energy
barrier compared to the dark states {[}see Fig. \ref{fig:Scheme_dynamics}a-II){]}.
However, at light-matter resonance, the R-ISC rate to the final polariton
state scales as $\frac{1}{N}$, given that the polariton is delocalized
across $N$ singlets and only one of them can undergo coupling to
a given triplet. Importantly, for $N\ge10^{4}$ the rate of transfer
to the polaritons falls below the cavity-free R-ISC rate for the entire
$(\Delta,\omega)$ range displayed (see Fig. \ref{fig:2Dplots} top
panels and \ref{fig:rates_bothcases}a). The conclusion is similar
for the slow bath case (b). While for small $N$ values, high efficiencies
are obtained (see Fig. \ref{fig:2Dplots}b-i and \ref{fig:2Dplots}b-ii),
the assistance of polariton modes decreases with increasing $N$.
This is because the ratio of rates (dark states $\rightarrow$ polariton)/(polariton$\rightarrow$
dark states) scales as $\frac{1}{N}$ given the delocalization of
the polariton state; this implies that the backward process polariton$\rightarrow$
dark states becomes more relevant as $N$ increases, an effect which
is detrimental to $\epsilon$. Finally, the decrease of photon character
in the MP for $\Delta<0$ {[}see Fig. \ref{fig:Scheme_dynamics}b-II){]}
together with a fast back transfer to the denser manifold of upper
dark states for larger $N$, explains the drop of $\epsilon$ at $\Delta<0$,
specially for $N>10^{4}$ (see Fig. \ref{fig:2Dplots} lower panels
and \ref{fig:rates_bothcases}b).

\begin{figure*}
\includegraphics[width=0.8\textwidth]{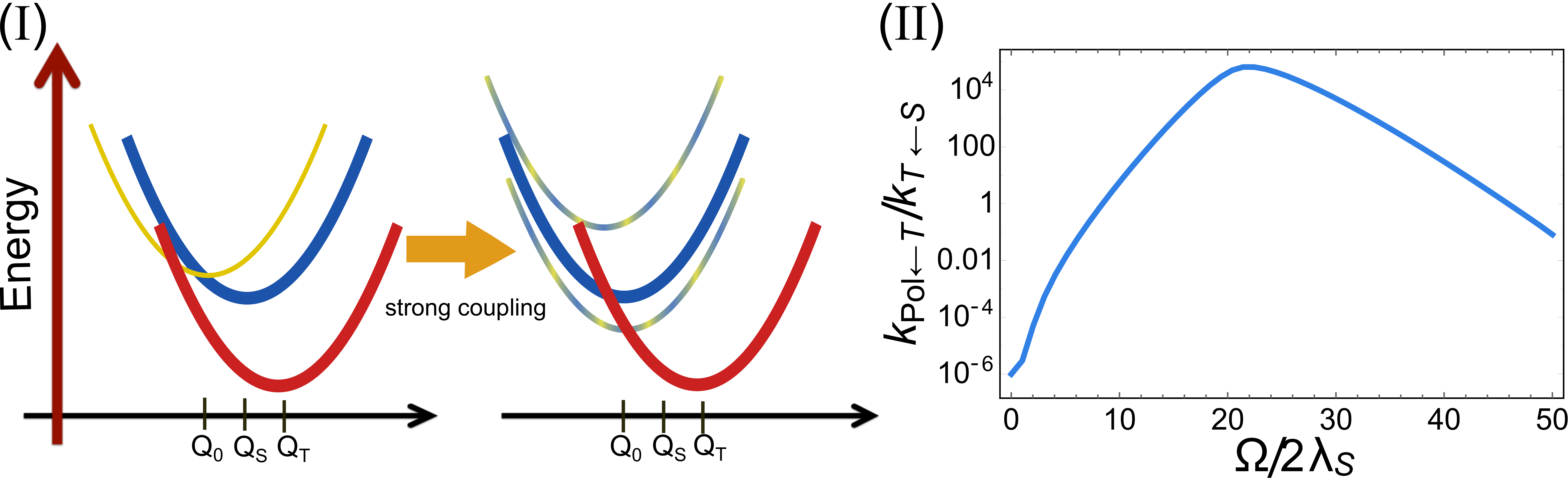}

\caption{Ideal case to maximize the rate of transfer from the triplet state
manifold (blue thick curve) to polariton states, with respect to the
cavity free triplet-to-singlet (R-ISC) transfer rate. (I) Energy diagrams
in the polaron frame. Left: a molecular emitter in which the singlet
(blue) and triplet (red) states have similar equilibrium nuclear configurations
and the dynamics between them falls within the Marcus inverted regime.
(I) Right: upon strong coupling to a photon mode and in the polaron
frame, the energy barrier between the triplet states and the lower
polariton (in this case, the lower polariton) quickly decreases with
Rabi splitting, giving rise to orders of magnitude increase in the
$k_{T\rightarrow\text{LP}}$ rate, thus avoiding the wash-out effect
of delocalization of the singlet of interest among $N$ molecules.
(II) Cavity mediated R-ISC enhancement $k_{T\to\text{LP}}/k_{T\rightarrow S}$.
Note that the cavity provides a realization of the various regimes
of Marcus theory, from normal to inverted, as a function of Rabi splitting.
The specific parameters for the vibrational bath are $a_{S}=0.5$,
$a_{T}=1$ and $\omega_{c}=0.01$ eV (see Theory section), which accounts
for a singlet reorganization energy $\lambda_{S}=0.01$ eV and $\lambda_{T}=0.02$
eV. We also assumed that the number of molecules per photon mode is
$N=10^{6}$, $\Delta G=(\epsilon_{S}-\lambda_{S})-(\epsilon_{T}-\lambda_{T})=0.2$
eV, where we recall that $\epsilon_{S}$ ($\epsilon_{T}$) is the
vertical singlet (triplet) transition energy; and the singlet-triplet
electronic coupling $V_{TS}=1\times10^{-5}$ eV.\label{fig:idealcase}}
\end{figure*}

Based on the discussion above, a pressing question is the characterization
of an optimal scenario that overcomes the deleterious effects of cavity
R-ISC rates as $N$ increases, the latter being as large as $10^{7}$
\cite{daskalakis2017}. Here, we propose a case where the energy tunability
of the polaritons introduces an increase of many orders of magnitude
relative to the cavity-free R-ISC rate, and this enhancement can outcompete
the suppression due to the low polariton DOS. The scenario is built
upon the fast bath case studied above, but assuming a sufficiently
large singlet-triplet energy gap and weak couplings of the electronic
states to the bath (small $a_{S}$ and $a_{T}$ parameters). Under
these conditions, the reduced dynamics of transfer in the polaron
frame in the cavity-free scenario corresponds to the inverted-Marcus
regime (for the singlet to triplet transition), where the population
of the triplet manifold needs to overcome an enormous energy barrier
to reach the singlet state (see Fig. \ref{fig:idealcase}). Upon confinement
to a microcavity, the emergent lower polariton could introduce a much
smaller energy barrier from the triplet states. Our results and the
precise parameters used in our calculations are shown in Fig. \ref{fig:idealcase},
which show that polariton-assistance introduces four orders of magnitude
enhancement relative to the bare R-ISC rate. This mechanism of energy
of activation reduction competing against delocalization is analogous
to the one presented in our recent work on vibrational polaritons
assisting thermally-activated reactions \cite{camposg}.

On the other hand, we note that the polariton-assisted R-ISC rate
is upper-bounded by $\frac{V_{ST}^{2}}{N}\sqrt{\frac{\pi\beta}{\lambda_{T-pol}}}$,
which must be higher than $k_{nrad}^{T}$ to lead to harvesting of
triplets. If we take $N=10^{6}$ , and use the parameters above for
the fast bath scenario ($k_{nrad}=2\times10^{-7}$ ps$^{-1}$), we
need a very small $\lambda_{T-pol}=10^{-17}$ eV. Therefore, in order
to obtain a sizable $\epsilon$, emitters with negligible couplings
to a vibrational environment are required (small reorganization energies
$\lambda_{ST}$ and $\lambda_{S}$); that is, poor R-ISC molecules
(\emph{i.e.}, the opposite of TADF materials) will obtain most benefit
from polaritonic effects. Whether this extreme possibility exists
will be subject of future work.

For the slow bath case, there is no possibility to enhance electroluminescence
for large $N$ because the associated R-ISC processes via polaritons
happen via one-phonon vibrational relaxation (Redfield) processes,
rather than multiphonon (Marcus) processes that exhibit exponential
sensitivity on energy barriers. Hence, the only option for enhancement
of electroluminescence for that case is to use samples with small
$N$ values.

\section{Conclusions\label{sec:Conclusions}}

In this article, we have developed a variational model for the prediction
of electroluminescence efficiency of molecular emitters interacting
with a photonic mode of a microcavity. The latter can describe the
emergent dynamics on a range of Rabi splittings that extend from the
weak to the the strong coupling regime.

Our calculations indicate that the main limitation to polariton-assist
the electroluminescence process is the delocalized character of polaritons:
the probability of a triplet state to R-ISC into the latter is strongly
suppressed by a $\frac{1}{N}$ factor, similarly to results discussed
in previous works \cite{Martinez2018,Du2018,Eizner2019}. There are
two approaches that could overcome the detrimental delocalization
effect and therefore introduce a polariton-enhanced electroluminescence
with respect to the cavity-free scenario: 1) for fast and slow bath
cases (\emph{i.e.} for chromophores with weak and strong singlet-triplet
mixing), as Figs. \ref{fig:2Dplots}a-i and \ref{fig:2Dplots}b-i
show, strong coupling with a small number of molecules per photonic
mode $N$ introduces significant electroluminescence yield enhancements
with respect to the bare molecular case. This regime should be feasible
in the context of plasmonic nanoparticles coated with organic dyes,
where $N$ can be more than 3 orders of magnitude smaller than those
needed in a microcavity to attain the strong coupling regime \cite{halas,Bisht2018,Munkhbat2018}.
2) The second approach only works for the fast bath case and consists
on tuning of polariton energies and taking advantage of the exponential
sensitivity of R-ISC rates with respect to energy barriers. Therefore,
a high thermal energy barrier for the R-ISC transition in the bare
molecular case translates into the possibility of its partial or total
reduction via a triplet to polariton R-ISC transition, thus potentially
outcompeting the delocalization effect, and introducing a net electroluminescence
enhancement. For a molecular emitter the previous requirement is equivalent
to an inverted Marcus regime for the singlet$\rightarrow$triplet
ISC transition. This mechanism is also reminiscent to a recently proposed
catalysis of thermally-activated reactions under vibrational strong
coupling, where a sufficient decrease in activation energy can outcompete
the large activation entropy of the dark states \cite{camposg}. An
inverted Marcus regime being the ideal scenario to introduce polariton
assistance has also been found in \cite{Martinez2018} for singlet
fission and in \cite{Semenov2019}, the latter in the context of a
single-molecule charge transfer process assisted by a photon mode.

On the other hand, for the scenario of a slow vibrational bath, the
efficiencies are largely determined by the energetic proximity of
the polariton and dark state resonances. The increase of Rabi splitting
in this case diminishes the rate of transition of the triplet to the
polariton state closer in energy, because the vibrational DOS that
can assist the transition decreases exponentially with the energy
gap betwen them. Therefore, for the slow phonon bath scenario considered
in this work, approach 1) is the only alternative to minimize the
polariton-delocalization effect on efficiencies.

\begin{figure}
\includegraphics[width=1\columnwidth]{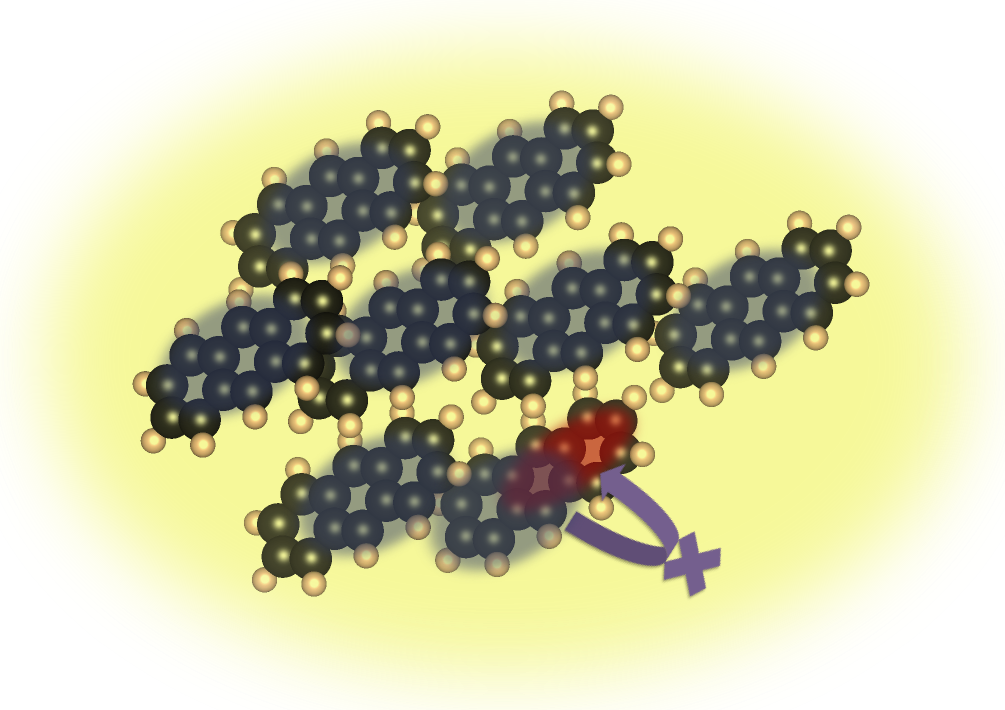}

\caption{Scheme that illustrates the delocalization of a singlet electronic
excitation (denoted as a faint blue on $N$ molecules) in a polariton
mode under strong coupling to a confined electromagnetic environment
(denoted as yellow). The localized nature of a triplet state (red)
precludes a fast transfer to the polariton modes, in view of the local
character of R-ISC and the dilution of probability of the R-ISC transition
among $N$ molecules. As elaborated in section V of this article,
this rate suppression can be overcome by 1) coupling fewer molecular
emitters to each photon mode while preserving strong light-matter
coupling (using e.g. plasmon nanoparticle systems); and/or ii) harnessing
strong coupling of molecules with singlet$\rightarrow$triplet transition
in the Marcus inverted regime, which feature very large R-ISC energy
barriers, such that introduction of a polariton channel may decrease
this barrier and possibly compete effectively against the many dark-state
channels.}

\end{figure}

\section*{Appendix A. Variational optimization of polaron transformation \label{sec:Appendix}}

\emph{Formalism.--- }According to the Feynman-Bogoliubov bound described
in section \ref{sec:Theory}, we are looking to variationally optimize
the set $\{\vec{X}^{(v)}\}$ of $v$-dependent parameter vectors $\vec{X}^{(v)}\equiv[f_{0}^{(v)},f_{1}^{(v)},f_{T}^{(v)},h]$
that minimize the free energy due to $\tilde{H}_{0}'$,
\begin{align*}
F_{0}'(\{\vec{X}^{(v)}\}) & =-\beta^{-1}\text{ln}\text{Tr}\left\{ \exp\Big[-\beta\tilde{H}_{0}'(\{\vec{X}^{(v)}\})\Big]\right\} \\
 & =-\beta^{-1}\text{ln}\Big[Z_{\tilde{H}_{0}}(\{\vec{X}^{(v)}\})\times Z_{H_{B}}\Big].
\end{align*}
Here, we have assumed $f_{1}^{(v)}=f_{2}^{(v)}=\dots=f_{N-1}^{(v)}\neq f_{0}^{(v)}$,
which is equivalent to making a distinction between the vibrational
deformation on a given molecular site and an indirect one induced
via the photon on the rest of the sites, $f_{1}^{(v)}\equiv f_{l\neq0}^{(v)}$.
Furthermore, 

\begin{subequations}\label{eq:Zs}
\begin{align}
Z_{\tilde{H}_{0}}(\{\vec{X}^{(v)}\}) & =\sum_{p=+,-}(N-1)e^{-\beta\tilde{\epsilon}_{p}(\{\vec{X}^{(v)}\})}\nonumber \\
 & +\sum_{q=\text{UP,MP,LP}}e^{-\beta\tilde{\epsilon}_{q}(\{\vec{X}^{(v)}\})},\label{eq:Z_H0}\\
Z_{H_{B}} & =N\prod_{v}(2\text{sinh}\frac{\beta\omega_{v}}{2})^{-1},\label{eq:Z_HB}
\end{align}

\end{subequations}\noindent are the corresponding polaritonic and
vibrational partition functions. Since $Z_{H_{B}}$ does not depend
on $\{\vec{X}^{(v)}\}$, we can minimize $F_{0}(\{\vec{X}^{(v)}\})=-\beta^{-1}\text{ln}Z_{\tilde{H}_{0}}(\{\vec{X}^{(v)}\})$
instead,

\begin{align}
\nabla_{\vec{X}^{(v)}}F_{0} & =\nabla_{\vec{Y}}F_{0}\cdot\nabla_{\vec{X}^{(v)}}\vec{Y}=0\hspace{1em}\forall v,\label{eq:free_ener_deriv}
\end{align}
where, using the chain rule, we have conveniently written $\vec{X}^{(v)}$
in terms of $\vec{Y}=[R_{S},R_{T},R_{ph},\eta_{ST},\eta_{S-ph}]$,
where {[}see Eqs. (\ref{renorm}){]}

\begin{subequations}

\begin{align}
R_{S} & =\sum_{v}\left(\sum_{l}\frac{f_{l}^{(v)2}}{\omega_{v}}-2g_{S}^{(v)}\frac{f_{0}^{(v)}}{\omega_{v}}\right),\label{eq:RS1}\\
R_{T} & =\sum_{v}\left(\frac{f_{T}^{(v)2}}{\omega_{v}}-2g_{T}^{(v)}\frac{f_{0}^{(v)}}{\omega_{v}}\right),\label{eq:RT1}\\
R_{ph} & =\sum_{v}N\frac{h^{(v)2}}{\omega_{v}}.\label{eq:Rph1}
\end{align}

\end{subequations}\noindent

The various terms in the $\nabla_{\vec{X}^{(v)}}\vec{Y}$ matrix are
provided in Table 1.

\begin{widetext}

\begin{table}[H]
\begin{centering}
\begin{tabular}{|c|c|c|c|c|c|}
\hline 
\multicolumn{6}{|c|}{TABLE 1. Derivatives $\frac{\partial y}{\partial x}$ where $y\in\{R_{S},R_{T},R_{ph},\eta_{ST},\eta_{S-ph}\}$
and $x\in\{f_{0}^{(v)},f_{1}^{(v)},f_{T}^{(v)},h\}$.\label{tab:Derivatives}}\tabularnewline
\hline 
$x\backslash y$ & $R_{S}$ & $R_{T}$ & $R_{ph}$ & $\eta_{ST}$ & $\eta_{S-ph}$\tabularnewline
\hline 
\hline 
$f_{0}^{(v)}$ & $\frac{2}{\omega_{v}}\left(f_{0}^{(v)}-g_{S}^{(v)}\right)$ & $0$ & $0$ & $-\frac{(f_{0}^{(v)}-f_{T}^{(v)})}{\omega_{v}^{2}}\coth(\beta\omega_{v}/2)\eta_{ST}$ & $-\frac{(f_{0}^{(v)}-h^{(v)})}{\omega_{v}^{2}}\coth(\beta\omega_{v}/2)\eta_{S-ph}$\tabularnewline
\hline 
$f_{l\neq0}^{(v)}$ & $\frac{2f_{1}^{(v)}}{\omega_{v}}$ & $0$ & $0$ & $-\frac{f_{1}^{(v)}}{\omega_{v}^{2}}\coth(\beta\omega_{v}/2)\eta_{ST}$ & $-\frac{(f_{1}^{(v)}-h^{(v)})}{\omega_{v}^{2}}\coth(\beta\omega_{v}/2)\eta_{S-ph}$\tabularnewline
\hline 
$f_{T}^{(v)}$ & $0$ & $\frac{2}{\omega_{v}}\left(f_{T}^{(v)}-g_{T}^{(v)}\right)$ & $0$ & $\frac{(f_{0}^{(v)}-f_{T}^{(v)})}{\omega_{v}^{2}}\coth(\beta\omega_{v}/2)\eta_{ST}$ & $0$\tabularnewline
\hline 
$h^{(v)}$ & $0$ & $0$ & $\frac{2Nh^{(v)}}{\omega_{v}}$ & $0$ & $\frac{(N-1)(f_{1}^{(v)}-h^{(v)})}{\omega_{v}^{2}}\coth(\beta\omega_{v}/2)\eta_{S-ph}$\tabularnewline
\hline 
\end{tabular}
\par\end{centering}
\end{table}

\end{widetext}

For every $v$, Eq. (\ref{eq:free_ener_deriv}) yields a set of four
equations in terms of the four unknown entries of $\vec{X}^{(v)}$.
Explicitly, for every $x\in\{f_{0}^{(v)},f_{1}^{(v)},f_{T}^{(v)},h\}$,
we solve $\partial_{x}F_{0}=0$ for $x$, obtaining:
\begin{widetext}
\begin{subequations}\label{syst_eqs}

\begin{align}
f_{0}^{(v)} & =\frac{2g_{S}^{(v)}\omega_{v}\frac{\partial F_{0}}{\partial R_{S}}-h^{(v)}\frac{\partial F_{0}}{\partial\eta_{S-ph}}\coth(\beta\omega_{v}/2)\eta_{S-ph}-\frac{\partial F_{0}}{\partial\eta_{ST}}f_{T}^{(v)}\coth(\beta\omega_{v}/2)\eta_{ST}}{2\frac{\partial F_{0}}{\partial R_{S}}\omega_{v}-\eta_{S-ph}\coth(\beta\omega_{v}/2)\frac{\partial F_{0}}{\partial\eta_{S-ph}}-\frac{\partial F_{0}}{\partial\eta_{ST}}\eta_{ST}\coth(\beta\omega_{v}/2)},\label{eq:sub1}\\
f_{1}^{(v)} & =\frac{h^{(v)}\frac{\partial F_{0}}{\partial\eta_{S-ph}}\coth(\beta\omega_{v}/2)\eta_{S-ph}}{-2\omega_{v}\frac{\partial F_{0}}{\partial R_{S}}+\eta_{S-ph}\coth(\beta\omega_{v}/2)\frac{\partial F_{0}}{\partial\eta_{S-ph}}+\frac{\partial F_{0}}{\partial\eta_{ST}}\coth(\beta\omega_{v}/2)\eta_{ST}},\\
f_{T}^{(v)} & =\frac{2\frac{\partial F_{0}}{\partial R_{T}}g_{T}^{(v)}\omega_{v}-\frac{\partial F_{0}}{\partial\eta_{ST}}f_{0}^{(v)}\eta_{ST}\coth(\beta\omega_{v}/2)}{2\frac{\partial F_{0}}{\partial R_{T}}\omega_{v}-\frac{\partial F_{0}}{\partial\eta_{ST}}\coth(\beta\omega_{v}/2)\eta_{ST}},\\
h^{(v)} & =\frac{-[f_{0}^{(v)}+(N-1)f_{1}^{(v)}]\frac{\partial F_{0}}{\partial\eta_{S-ph}}\coth(\beta\omega_{v}/2)\eta_{S-ph}}{2\omega_{v}N\frac{\partial F_{0}}{\partial R_{ph}}-N\frac{\partial F_{0}}{\partial\eta_{S-ph}}\coth(\beta\omega_{v}/2)\eta_{S-ph}}.
\end{align}

\end{subequations}\noindent Analytical expressions for the derivatives
$\partial_{y}F_{0}$ can be found in Appendix B. Next, we take the
continuum limit of Eq. (\ref{syst_eqs}) by making the substitutions,
$f_{l}^{(v)}\rightarrow f_{l}(\omega)$, $f_{T}^{(v)}\rightarrow f_{T}(\omega)$,
$h^{(v)}\rightarrow h(\omega)$, and introducing the vibrational DOS
$\mathcal{D}(\omega)=\omega\omega_{c}^{-2}e^{-\omega/\omega_{c}}$,
upon which the bath spectral densities become $J_{i}(\omega)=a_{i}\frac{\omega^{3}}{\omega_{c}^{2}}e^{-\omega/\omega_{c}}=|g_{i}(\omega)|^{2}\mathcal{D}(\omega)$,
$i=S,T$, where $|g_{i}(\omega)|^{2}=a_{i}\omega^{2}$ is the frequency-dependent
vibronic-coupling intensity for each excitation. Since the spectral
densities encode information on the geometries of the $S$ and $T$
states with respect to the ground molecular state, it is clear that
they allow for the calculation of the spectral density for the singlet-triplet
transition as $J_{ST}(\omega)=a_{ST}\frac{\omega^{3}}{\omega_{c}^{2}}e^{-\omega/\omega_{c}}$,
$a_{ST}=(a_{T}^{1/2}\mp a_{S}^{1/2})^{2}$, as well as the corresponding
reorganization energy $\lambda_{ST}=\int_{0}^{\infty}\frac{J_{ST}(\omega)}{\omega}d\omega$.
Here, for each mode of frequency $\omega$, the minus sign should
be chosen if the nuclear geometries of the $S$ and $T$ states shift
in the same direction with respect to the ground molecular state,
and the plus sign otherwise; we hereby assume the former, in light
of geometries calculated for typical TADF molecules \cite{Peng2017}.
With these considerations, Eq. (\ref{syst_eqs}) can be manipulated
to write $f_{1}$, $h$, and $f_{T}$ in terms of $f_{0}$ and to
solve for $f_{0}$,

\begin{subequations}\label{eq:final_eqs} 
\begin{align}
f_{0}(\omega) & =\frac{\mathcal{R}(\omega)}{\mathcal{G}(\omega)+\frac{\partial F{}_{0}}{\partial\eta_{S-ph}}\eta_{S-ph}\coth(\beta\omega/2)(-2\frac{\partial F{}_{0}}{\partial R_{T}}\omega+\frac{\partial F{}_{0}}{\partial\eta_{ST}}\eta_{ST}\coth(\beta\omega/2))(1+(N-1)\mathcal{T}(\omega))\chi(\omega)},\label{eq:final_1}\\
f_{1}(\omega) & =\mathcal{T}(\omega)f_{0}(\omega),\label{eq:final_2}\\
h(\omega) & =-\Big[1+(N-1)\mathcal{T}(\omega)\Big]\chi(\omega)f_{0}(\omega),\label{eq:final_3}\\
f_{T}(\omega) & =\frac{2\frac{\partial F_{0}}{\partial R_{T}}g_{T}(\omega)\omega-\frac{\partial F_{0}}{\partial\eta_{ST}}\eta_{ST}\coth(\beta\omega/2)f_{0}(\omega)}{2\frac{\partial F_{0}}{\partial R_{T}}\omega-\frac{\partial F_{0}}{\partial\eta_{ST}}\coth(\beta\omega/2)\eta_{ST}}.\label{eq:final_4}
\end{align}

\end{subequations} These expressions depend on the following auxiliary
functions,

\begin{subequations}\label{eq:auxiliary} 
\begin{align}
\mathcal{R}(\omega) & =2\omega\left[2\frac{\partial F{}_{0}}{\partial R_{S}}\frac{\partial F{}_{0}}{\partial R_{T}}g_{S}(\omega)\omega-\frac{\partial F{}_{0}}{\partial\eta_{ST}}\left(\frac{\partial F{}_{0}}{\partial R_{S}}g_{S}(\omega)+\frac{\partial F{}_{0}}{\partial R_{T}}g_{T}(\omega)\right)\eta_{ST}\coth(\beta\omega/2)\right],\label{eq:aux_1}\\
\mathcal{G}(\omega) & =4\frac{\partial F{}_{0}}{\partial R_{S}}\frac{\partial F{}_{0}}{\partial R_{T}}\omega^{2}\nonumber \\
 & \hspace{1em}-2\left[\frac{\partial F{}_{0}}{\partial\eta_{S-ph}}\frac{\partial F{}_{0}}{\partial R_{T}}\eta_{S-ph}+\frac{\partial F{}_{0}}{\partial\eta_{ST}}\left(\frac{\partial F{}_{0}}{\partial R_{S}}+\frac{\partial F{}_{0}}{\partial R_{T}}\right)\eta_{ST}\right]\omega\coth(\beta\omega/2)\nonumber \\
 & \hspace{1em}+\frac{\partial F{}_{0}}{\partial\eta_{S-ph}}\frac{\partial F{}_{0}}{\partial\eta_{ST}}\eta_{S-ph}\eta_{ST}\coth(\beta\omega/2)^{2},\label{eq:aux_2}\\
\mathcal{T}(\omega) & =\frac{\left(\frac{\partial F{}_{0}}{\partial\eta_{S-ph}}\right)^{2}\eta_{S-ph}^{2}\coth(\beta\omega/2)^{2}}{\mathcal{W}(\omega)},\label{eq:aux_3}\\
\mathcal{W}(\omega) & =4N\frac{\partial F{}_{0}}{\partial R_{S}}\frac{\partial F{}_{0}}{\partial R_{ph}}\omega^{2}\nonumber \\
 & \hspace{1em}-2\left[\frac{\partial F{}_{0}}{\partial\eta_{S-ph}}\eta_{S-ph}\left(N\frac{\partial F{}_{0}}{\partial R_{ph}}+N\frac{\partial F{}_{0}}{\partial R_{S}}\right)+N\frac{\partial F{}_{0}}{\partial\eta_{ST}}\frac{\partial F{}_{0}}{\partial R_{ph}}\eta_{ST}\right]\omega\coth(\beta\omega/2)\nonumber \\
 & \hspace{1em}+\frac{\partial F{}_{0}}{\partial\eta_{S-ph}}\eta_{S-ph}\left(\frac{\partial F{}_{0}}{\partial\eta_{S-ph}}\eta_{S-ph}+N\frac{\partial F{}_{0}}{\partial\eta_{ST}}\eta_{ST}\right)\coth(\beta\omega/2)^{2},\label{eq:aux_4}\\
\mathcal{\chi}(\omega) & =\frac{\frac{\partial F{}_{0}}{\partial\eta_{S-ph}}\coth(\beta\omega/2)\eta_{S-ph}}{2\omega N\frac{\partial F{}_{0}}{\partial R_{ph}}-N\frac{\partial F{}_{0}}{\partial\eta_{phot}}\coth(\beta\omega/2)\eta_{S-ph}}.\label{eq:aux_5}
\end{align}
\end{subequations} Eq. (\ref{eq:final_eqs}) expresses $\vec{X}(\omega)=[f_{0}(\omega),f_{1}(\omega),f_{T}(\omega),h(\omega)]$
in terms of the known distributions $g_{i}(\omega)$, the renormalization
parameters $\eta_{S-ph},\eta_{ST}$ and the derivatives $\frac{\partial F_{0}}{\partial y}$,
which permit a self-consistent solution scheme as explained below. 

Finally, the renormalization parameters can also be expressed according
to,

\begin{subequations}\label{eq:renorm}

\begin{align}
R_{S} & =-2\int_{0}^{\infty}d\omega g_{S}(\omega)\frac{f_{0}(\omega)}{\omega}\mathcal{D}(\omega)\nonumber \\
 & \hspace{1em}+(N-1)\int_{0}^{\infty}d\omega\frac{f_{1}(\omega)^{2}}{\omega}\mathcal{D}(\omega)\nonumber \\
 & \hspace{1em}+\int_{0}^{\infty}d\omega\frac{f_{0}(\omega)^{2}}{\omega}\mathcal{D}(\omega)\nonumber \\
 & =\int_{0}^{\infty}d\omega\left[-2g_{S}(\omega)+f_{0}(\omega)\right]\frac{f_{0}(\omega)}{\omega}\mathcal{D}(\omega)\nonumber \\
 & \hspace{1em}+(N-1)\int_{0}^{\infty}d\omega\frac{|\mathcal{T}(\omega)f_{0}(\omega)|^{2}}{\omega}\mathcal{D}(\omega),\label{eq:R_S}\\
R_{T} & =-2\int_{0}^{\infty}d\omega g_{T}(\omega)\frac{f_{0}^{T}(\omega)}{\omega}\mathcal{D}(\omega)\nonumber \\
 & \hspace{1em}+\int_{0}^{\infty}\frac{|f_{0}^{T}(\omega)|^{2}}{\omega}\mathcal{D}(\omega)d\omega,\label{eq:R_T}\\
R_{ph} & =N\int_{0}^{\infty}\frac{h(\omega)^{2}}{\omega}\mathcal{D}(\omega)d\omega\nonumber \\
 & =N\int_{0}^{\infty}\frac{f_{0}(\omega)^{2}}{\omega}\Big[1+(N-1)\mathcal{T}(\omega)\Big]^{2}\chi(\omega)^{2}\mathcal{D}(\omega)d\omega\label{eq:R_ph}\\
\eta_{S-ph} & =\exp\Bigg\{-\frac{1}{2}\int_{0}^{\infty}d\omega\Bigg[\left(\frac{f_{0}(\omega)-h(\omega)}{\omega}\right)^{2}+(N-1)\left(\frac{f_{1}(\omega)-h(\omega)}{\omega}\right)^{2}\Bigg]\mathcal{D}(\omega)\coth(\beta\omega/2)\Bigg\}\nonumber \\
 & =\exp\Bigg[-\frac{1}{2}\int_{0}^{\infty}d\omega f_{0}(\omega)^{2}\Bigg(\Bigg\{\frac{1+[1+(N-1)\mathcal{T}(\omega)]\chi(\omega)}{\omega}\Bigg\}^{2}+(N-1)\Bigg\{\frac{\mathcal{T}(\omega)+[1+(N-1)\mathcal{T}(\omega)]\chi(\omega)}{\omega}\Bigg\}^{2}\Bigg)\nonumber \\
 & \hspace{1em}\times\mathcal{D}(\omega)\coth(\beta\omega/2)\Bigg],\label{eq:Nphot}\\
\eta_{ST} & =\exp\left[-\frac{1}{2}\int_{0}^{\infty}d\omega\left(\frac{f_{0}(\omega)-f_{T}(\omega)}{\omega}\right)^{2}\coth(\beta\omega/2)\mathcal{D}(\omega)-\frac{1}{2}(N-1)\int_{0}^{\infty}d\omega\frac{|f_{1}(\omega)|^{2}}{\omega^{2}}\coth(\beta\omega/2)\mathcal{D}(\omega)\right]\nonumber \\
 & =\exp\left[-\frac{1}{2}\int_{0}^{\infty}d\omega\left(\frac{f_{0}(\omega)-f_{T}(\omega)}{\omega}\right)^{2}\coth(\beta\omega/2)\mathcal{D}(\omega)-\frac{N-1}{2}\int_{0}^{\infty}d\omega\left(\frac{\mathcal{T}(\omega)f_{0}(\omega)}{\omega}\right)^{2}\coth(\beta\omega/2)\mathcal{D}(\omega)\right].\label{eq:NTS}
\end{align}
\end{subequations}

\emph{Analytical evaluation of $\nabla_{\vec{Y}}F_{0}$.---} Let
$Z_{\tilde{H}_{0,k=0}}=\sum_{j=\text{UP,MP,LP}}e^{-\beta\tilde{\epsilon}_{j}}$,
so that Eq. (\ref{eq:Z_H0}) can be rewritten as $Z_{\tilde{H}_{0}}=Z_{\tilde{H}_{0,k=0}}+\sum_{p=+,-}(N-1)e^{-\beta\tilde{\epsilon}_{p}}$.
Then, the various entries of $\nabla_{\vec{Y}}F_{0}$ can be obtained
from,

\begin{widetext}

\begin{subequations}\label{eq:derivatives}

\begin{align}
-\beta Z_{\tilde{H}_{0}}\frac{\partial F_{0}}{\partial R_{S}} & =\frac{\partial Z_{\tilde{H}_{0,k=0}}}{\partial R_{S}}-\frac{\beta(N-1)}{2}\sum_{p=\pm}e^{-\beta\tilde{\epsilon}_{p}}\left(1+p\frac{\tilde{\epsilon}_{S}-\tilde{\epsilon}_{T}}{\sqrt{(\tilde{\epsilon}_{S}-\tilde{\epsilon}_{T})^{2}+4\tilde{V}_{ST}^{2}}}\right),\label{eq:renorm1}\\
-\beta Z_{\tilde{H}_{0}}\frac{\partial F_{0}}{\partial R_{T}} & =\frac{\partial Z_{\tilde{H}_{0,k=0}}}{\partial R_{T}}-\frac{\beta(N-1)}{2}\sum_{p=\pm}e^{-\beta\tilde{\epsilon}_{p}}\left(1-p\frac{\tilde{\epsilon}_{S}-\tilde{\epsilon}_{T}}{\sqrt{(\tilde{\epsilon}_{S}-\tilde{\epsilon}_{T})^{2}+4\tilde{V}_{ST}^{2}}}\right),\label{eq:renorm2}\\
-\beta Z_{\tilde{H}_{0}}\frac{\partial F_{0}}{\partial R_{ph}} & =\frac{\partial Z_{\tilde{H}_{0,k=0}}}{\partial R_{ph}},\label{eq:renorm3}\\
-\beta Z_{\tilde{H}_{0}}\frac{\partial F_{0}}{\partial\eta_{TS}} & =\frac{\partial Z_{\tilde{H}_{0,k=0}}}{\partial\eta_{TS}}+\frac{2(N-1)\eta_{ST}V_{ST}^{2}\beta}{\sqrt{(\tilde{\epsilon}_{S}-\tilde{\epsilon}_{T})^{2}+4\tilde{V}_{ST}^{2}}}\Big(e^{-\beta\epsilon_{-}}e^{-\beta\epsilon_{+}}\Big),\label{eq:renorm4}\\
-\beta Z_{\tilde{H}_{0}}\frac{\partial F_{0}}{\partial\eta_{ph}} & =\frac{\partial Z_{\tilde{H}_{0,k=0}}}{\partial\eta_{ph}}.\label{eq:renorm5}
\end{align}

\end{subequations}

\end{widetext}

The expressions above depend on 
\begin{equation}
\partial_{y}Z_{\tilde{H}_{0,k=0}}=-\beta\sum_{j=\text{UP,MP,LP}}\partial_{y}\tilde{\epsilon}_{j}e^{-\beta\tilde{\epsilon}_{j}}.\label{eq:der_wrt_y}
\end{equation}
The required derivatives $\partial_{y}\tilde{\epsilon}_{j}$ can be
obtained as follows: the secular equation associated to the matrix
$\tilde{H}_{k=0}$ ($\tilde{g}=g\eta_{S-ph}$) is

\begin{widetext}
\begin{align*}
\tilde{\epsilon}_{j}^{3}-\tilde{\epsilon}_{j}^{2}(\tilde{\omega}_{ph}+\tilde{\epsilon}_{S}+\tilde{\epsilon}_{T}) & +\tilde{\epsilon}_{j}(\tilde{\epsilon}_{S}\tilde{\epsilon}_{T}+\tilde{\omega}_{ph}\tilde{\epsilon}_{T}+\tilde{\omega}_{ph}\tilde{\epsilon}_{S}-N\tilde{g}^{2}-\tilde{V}_{ST}^{2})-\tilde{\epsilon}_{S}\tilde{\epsilon}_{T}\tilde{\omega}_{ph}+N\tilde{g}^{2}\tilde{\epsilon}_{T}+\tilde{V}_{ST}^{2}\tilde{\omega}_{ph}=0
\end{align*}

\end{widetext}

from which

\begin{widetext}\begin{subequations}\label{eq:final_derivatives}
\begin{align}
\zeta_{j}\frac{\partial\tilde{\epsilon}_{j}}{\partial R_{T}} & =\tilde{\epsilon}_{j}^{2}-\tilde{\epsilon}_{j}(\tilde{\epsilon}_{S}+\tilde{\omega}_{ph})+\tilde{\epsilon}_{S}\tilde{\omega}_{ph}-N\tilde{g}^{2},\label{eq:der1}\\
\zeta_{j}\frac{\partial\tilde{\epsilon}_{j}}{\partial R_{S}} & =\tilde{\epsilon}_{T}\tilde{\omega}_{ph}-(\tilde{\epsilon}_{T}+\tilde{\omega}_{ph})\tilde{\epsilon}_{j}+\tilde{\epsilon}_{j}^{2},\label{eq:der2}\\
\zeta_{j}\frac{\partial\tilde{\epsilon}_{j}}{\partial\eta_{TS}} & =2\tilde{V}_{ST}V_{ST}\left(\tilde{\epsilon}_{j}-\tilde{\omega}_{ph}\right),\label{eq:der3}\\
\zeta_{j}\frac{\partial\tilde{\epsilon}_{j}}{\partial R_{ph}} & =\tilde{\epsilon}_{S}\tilde{\epsilon}_{T}-\tilde{V}_{ST}^{2}-\tilde{\epsilon}_{S}\lambda-\tilde{\epsilon}_{T}\lambda+\lambda^{2},\label{eq:der4}\\
\zeta_{j}\frac{\partial\tilde{\epsilon}_{j}}{\partial\eta_{S-ph}} & =2N\tilde{g}g\left(\tilde{\epsilon}_{j}-\tilde{\epsilon}_{T}\right),\label{eq:der5}
\end{align}

\end{subequations}\end{widetext}\noindent where $\zeta_{j}=3\tilde{\epsilon}_{j}^{2}-2\tilde{\epsilon}_{j}(\tilde{\omega}_{ph}+\tilde{\epsilon}_{S}+\tilde{\epsilon}_{T})+\tilde{\epsilon}_{S}\tilde{\epsilon}_{T}+\tilde{\omega}_{ph}\tilde{\epsilon}_{T}+\tilde{\omega}_{ph}\tilde{\epsilon}_{S}-N\tilde{g}^{2}-\tilde{V}_{ST}^{2}$.
Thus, numerical diagonalization of $\tilde{H}_{0,k=0}$ (see Eq. (\ref{eq:H_0_k_0}))
to obtain $\{\epsilon_{j}\}$ and use of Eq. (\ref{eq:final_derivatives})
yields a robust evaluation of Eq. (\ref{eq:der_wrt_y}).

\emph{Numerical implementation.---} Fig. \ref{fig:flowchart} summarizes
the numerical implementation of the variational optimization using
the formalism outlined above. 
\end{widetext}

\begin{figure*}
\begin{centering}
\includegraphics[scale=0.6]{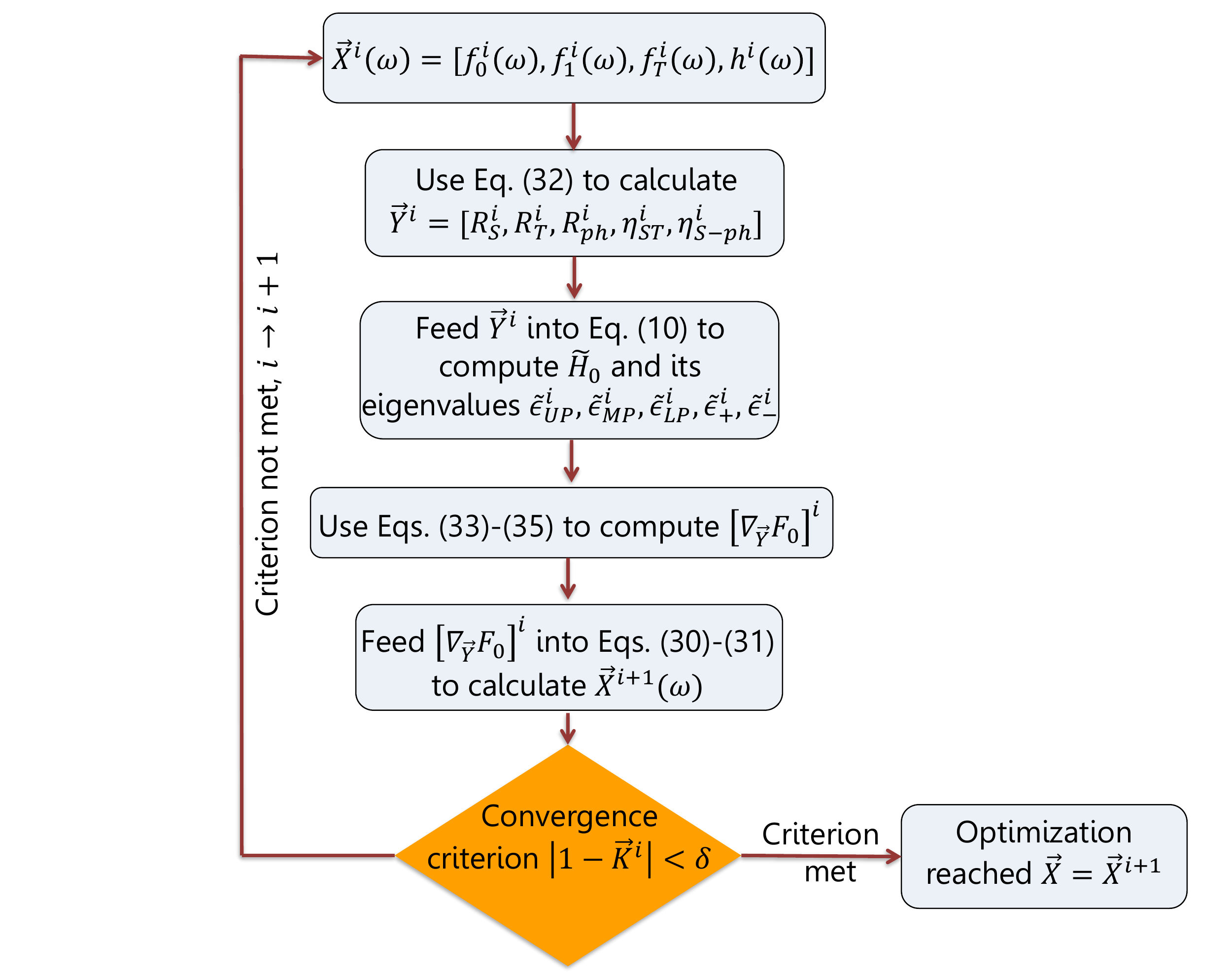}
\par\end{centering}
\caption{Flowchart that illustrates the self-consistent numerical optimization
of $\vec{X}(\omega)=[f_{0}(\omega),f_{1}(\omega),f_{T}(\omega),h(\omega)]$.
The $i$th iteration of the algorithm starts with $\vec{X}^{i}(\omega)=[f_{0}^{i}(\omega),f_{1}^{i}(\omega),f_{T}^{i}(\omega),h^{i}(\omega)]$
which at the beginning $(i=0)$, is chosen as the guess $\vec{X}^{0}(\omega)=[g_{S}(\omega),0,g_{T}(\omega),0]$
(corresponding to a full-polaron transformation ansatz). $\vec{X}^{i}(\omega)$
is subsequently used in Eq. (\ref{eq:renorm}) for the calculation
of $\vec{Y}^{i}=[R_{S}^{i},R_{T}^{i},R_{ph}^{i},\eta_{ST}^{i},\eta_{S-ph}^{i}]$,
which allows to build and numerically solve the renormalized Hamiltonian
(\ref{eq:zeroth}). The eigenvalues of the latter ($\tilde{\epsilon}_{UP}^{i}$,$\tilde{\epsilon}_{MP}^{i}$,$\tilde{\epsilon}_{LP}^{i}$,$\tilde{\epsilon}_{+}^{i},\tilde{\epsilon}_{-}^{i}$),
in conjuction with Eqs. (\ref{eq:derivatives})--(\ref{eq:final_derivatives})
to calculate $[\nabla_{\vec{Y}}F_{0}]^{i}$ which, via Eqs. (\ref{eq:final_eqs})--(\ref{eq:auxiliary}),
yield the guess $\vec{X}^{i+1}(\omega)$ to be used in the next iteration.
This process is carried out until the convergence criterion is reached,
which we define as $|\mathbf{1}-\vec{K}^{i}|<\delta$, where $\mathbf{1}=[1,1,1,1,1]$,
$\vec{K}^{i}=\Bigg[\frac{R_{S}^{i+1}}{R_{S}^{i}},\frac{R_{T}^{i+1}}{R_{T}^{i}},\frac{R_{ph}^{i+1}}{R_{ph}^{i}},\frac{\eta_{ST}^{i+1}}{\eta_{ST}^{i}},\frac{\eta_{S-ph}^{i+1}}{\eta_{S-ph}^{i}}\Bigg]$,
and $\delta=10^{-6}$. \label{fig:flowchart}}
\end{figure*}

\section*{Appendix B. Evaluation of rates \label{sec:Appendix-B}}

The rate of population transfer from eigenstate $m$ to $n$ of $\tilde{H}_{0}$
(see Eq. \ref{eq:zeroth}) is mediated by the residual interaction
$\tilde{H}_{I}$ in Eq. (\ref{res_int}), and can be calculated in
terms of the corresponding autocorrelation function,

\begin{widetext}
\begin{widetext}
\begin{align}
k_{m\rightarrow n} & =\int_{-\infty}^{\infty}\langle\tilde{H}_{I}^{(m,n)}(\tau)\tilde{H}_{I}^{(m,n)}(0)^{*}\rangle d\tau\nonumber \\
 & =\int_{-\infty}^{\infty}\big\langle\left[\tilde{V}_{ST}^{(m,n)}(\tau)+\tilde{V}_{S-ph}^{(m,n)}(\tau)+\tilde{V}_{S}^{(m,n)}(\tau)+\tilde{V}_{T}^{(m,n)}(\tau)+\tilde{V}_{ph}^{(m,n)}(\tau)\right]\nonumber \\
 & \hspace{1em}\times\left[\tilde{V}_{ST}^{(m,n)*}(0)+\tilde{V}_{S-ph}^{(m,n)*}(0)+\tilde{V}_{S}^{(m,n)*}(0)+\tilde{V}_{T}^{(m,n)*}(0)+\tilde{V}_{ph}^{(m,n)*}(0)\right]\big\rangle d\tau,\label{FGR_gral}
\end{align}

\end{widetext}\noindent where $\tilde{V}_{x}^{(m,n)}(\tau)=\langle m|e^{i\tilde{H}_{S}\tau}\tilde{V}_{x}e^{-i\tilde{H}_{S}\tau}|n\rangle=e^{i(\omega_{m}-\omega_{n})\tau}\langle m|e^{iH_{B}\tau}\tilde{V}_{x}e^{-iH_{B}\tau}|n\rangle$
for $x=ST,\,S-ph,\,S,\,T,\,ph$. 

From Eq. (\ref{FGR_gral}), we notice the emergence of auto- and cross-correlation
terms which fall into one of three types, and are summarized in Table
\ref{tab:correlation}. Correlations of similar nature have been explored
in previous works using a variational approach for energy transfer
dynamics in chromophoric arrays \cite{Mccutcheon2011b,Pollock2013}.
In Table \ref{tab:correlation} we also show the specific forms of
the correlation functions in time domain in the localized basis, upon
which the rates in Eq. (\ref{FGR_gral}) can be calculated. Notice
that we can further classify the correlations according to whether
they emerge from averaging over the vibrational Hamiltonians of same
or different electronic sites (see Table \ref{tab:correlation}).%%%%TABLE starts here
%\begin{figure*}
\renewcommand\thetable{2}
\begin{table*}
\begin{tabular}{|c||c||c||c|c|}
\hline 
\multicolumn{5}{|c|}{%\begin{tabular*}{1\linewidth}{@{\extracolsep{\fill}}|c||c||c||c|c}
Summary of relevant correlation functions for dynamics calculations}\tabularnewline
\hline 
\hline 
\multicolumn{5}{|c|}{Type: exponential-exponential}\tabularnewline
\hline 
\multicolumn{4}{|c|}{Same site} & Different sites\tabularnewline
\hline 
\multicolumn{4}{|c|}{$\begin{aligned} & \langle e^{\hat{D}_{ST}^{(n)}(\tau)}e^{\lambda\hat{D}_{ST}^{(n)}(0)}\rangle=\eta_{ST}^{2}\exp\big[Q(\lambda|\Delta_{ST}^{(0)}(\omega)|^{2}\\
 & +\lambda(N-1)|f_{1}|^{2}(\omega),\tau)\big]
\end{aligned}
$} & $\begin{aligned} & \langle e^{\hat{D}_{ST}^{(n)}(\tau)}e^{\lambda\hat{D}_{ST}^{(n)}(0)}\rangle=\eta_{ST}^{2}\exp\big[Q(\lambda|\Delta_{ST}^{(0)}(\omega)|^{2}\\
 & +\lambda(N-1)|f_{1}|^{2}(\omega),\tau)\big]
\end{aligned}
$ \tabularnewline
\hline 
\multicolumn{4}{|c|}{$\begin{aligned} & \langle e^{\hat{D}_{S-ph}^{(n)}(\tau)}e^{\lambda\hat{D}_{S-ph}^{(n)}}\rangle=\eta_{S-ph}^{2}\exp\bigg[Q(\lambda|\Delta_{S-ph}^{(0)}(\omega)|^{2}\\
 & +\lambda(N-1)|f_{1}(\omega)-h(\omega)|^{2},\tau)\bigg]
\end{aligned}
$} & $\begin{aligned} & \langle e^{\hat{D}_{S-ph}^{(n)}(\tau)}e^{\lambda\hat{D}_{S-ph}^{(m)}}\rangle=\eta_{S-ph}^{2}\exp\bigg[Q(2\lambda\Delta_{S-ph}^{(0)}(\omega)\\
 & \times(f_{1}(\omega)-h(\omega))+\lambda(N-2)|f_{1}(\omega)-h(\omega)|^{2},\tau)\bigg]
\end{aligned}
$ \tabularnewline
\hline 
\multicolumn{4}{|c|}{$\begin{aligned} & \langle e^{\hat{D}_{ST}^{(n)}(\tau)}e^{\lambda\hat{D}_{S-ph}^{(n)}(0)}\rangle=\eta_{ST}\eta_{S-ph}\exp\bigg[Q(\lambda|\Delta_{S-ph}^{(0)}(\omega)|^{2}+\\
 & +\lambda(N-1)|f_{1}(\omega)-h(\omega)|^{2},\tau)\bigg]
\end{aligned}
$} & $\begin{aligned} & \langle e^{\hat{D}_{ST}^{(n)}(\tau)}e^{\lambda\hat{D}_{S-ph}^{(m)}(0)}\rangle=\eta_{ST}\eta_{S-ph}\\
 & \exp\bigg\{ Q\bigg[\lambda\bigg(\Delta_{S-ph}^{(0)}(\omega)f_{1}(\omega)\\
 & +(f_{1}(\omega)-h(\omega))\Delta_{ST}^{(0)}(\omega)\\
 & +(N-2)f_{1}(\omega)(f_{1}(\omega)-h(\omega))\bigg),\tau\bigg\}
\end{aligned}
$ \tabularnewline
\hline 
\multicolumn{5}{|c|}{Type: exponential-linear}\tabularnewline
\hline 
\multicolumn{4}{|c|}{$\begin{aligned} & \langle e^{\lambda\hat{D}_{ST}^{(n)}(\tau)}\hat{D}_{S}^{(n)}(0)\rangle=\int_{0}^{\infty}d\omega\mathcal{D}(\omega)\lambda\big[\frac{(g_{S}(\omega)-f_{0}(\omega))\Delta_{ST}^{(0)}(\omega)}{\omega}+\\
 & (N-1)\frac{|f_{1}(\omega)|^{2}}{\omega}\big]\times\left[\overline{n}(\omega)e^{i\omega\tau}-(\overline{n}(\omega)+1)e^{-i\omega\tau}\right]\eta_{ST}
\end{aligned}
$} & $\begin{aligned} & \langle e^{\lambda\hat{D}_{ST}^{(n)}(\tau)}\hat{D}_{S}^{(m)}(0)\rangle=\int_{0}^{\infty}d\omega\mathcal{D}(\omega)\lambda\bigg[(g_{S}(\omega)-f_{0}(\omega)\\
 & -(N-2)f_{1}(\omega))\frac{f_{1}(\omega)}{\omega}-f_{1}(\omega)\frac{\Delta_{ST}^{(0)}(\omega)}{\omega}\bigg]\\
 & \times\left[\overline{n}(\omega)e^{i\omega\tau}-(\overline{n}(\omega)+1)e^{-i\omega\tau}\right]\eta_{ST}
\end{aligned}
$ \tabularnewline
\hline 
\multicolumn{4}{|c|}{$\begin{aligned} & \langle e^{\lambda\hat{D}_{ST}^{(n)}(\tau)}\hat{D}_{T}^{(n)}(0)\rangle=\int_{0}^{\infty}d\omega\mathcal{D}(\omega)\lambda\frac{(g_{T}(\omega)-f_{T}(\omega))\Delta_{ST}^{(0)}(\omega)}{\omega}\\
 & \times\left[\overline{n}(\omega)e^{i\omega\tau}-(\overline{n}(\omega)+1)e^{-i\omega\tau}\right]\eta_{ST}d\omega
\end{aligned}
$} & $\begin{aligned} & \langle e^{\lambda\hat{D}_{ST}^{(n)}(\tau)}\hat{D}_{T}^{(m)}(0)\rangle=\int_{0}^{\infty}d\omega\mathcal{D}(\omega)\lambda\frac{(g_{T}(\omega)-f_{T}(\omega))f_{1}(\omega)}{\omega}\\
 & \times\left[\overline{n}(\omega)e^{i\omega\tau}-(\overline{n}(\omega)+1)e^{-i\omega\tau}\right]\eta_{ST}
\end{aligned}
$ \tabularnewline
\hline 
\multicolumn{4}{|c|}{$\begin{aligned} & \langle e^{\lambda\hat{D}_{S-ph}^{(n)}(\tau)}\hat{D}_{S}^{(n)}(0)\rangle=\int_{0}^{\infty}d\omega\mathcal{D}(\omega)\lambda\bigg[\frac{(g_{S}(\omega)-f_{0}(\omega))\Delta_{S-ph}^{(0)}(\omega)}{\omega}\\
 & -(N-1)\frac{f_{1}(\omega)\Delta_{S-ph}^{(1)}(\omega)}{\omega}\bigg]\\
 & \times\left[\overline{n}(\omega)e^{i\omega\tau}-(\overline{n}(\omega)+1)e^{-i\omega\tau}\right]\eta_{S-ph}
\end{aligned}
$} & $\begin{aligned} & \langle e^{\lambda\hat{D}_{S-ph}^{(n)}(\tau)}\hat{D}_{S}^{(m)}(0)\rangle=\int_{0}^{\infty}d\omega\mathcal{D}(\omega)\lambda\bigg[2\frac{(g_{S}(\omega)-f_{0}(\omega))}{\omega}\\
 & \times\Delta_{S-ph}^{(1)}(\omega)-(N-2)\frac{f_{1}(\omega)\Delta_{S-ph}^{(1)}(\omega)}{\omega}\bigg]\\
 & \times\left[\overline{n}(\omega)e^{i\omega\tau}-(\overline{n}(\omega)+1)e^{-i\omega\tau}\right]\eta_{S-ph}
\end{aligned}
$ \tabularnewline
\hline 
\multicolumn{5}{|c|}{Type: linear-linear}\tabularnewline
\hline 
\multicolumn{4}{|c|}{$\begin{aligned} & \langle\hat{D}_{S}^{(n)}(\tau)\hat{D}_{S}^{(n)}(0)\rangle=\int_{0}^{\infty}d\omega\mathcal{D}(\omega)\big[|g_{S}(\omega)-f_{0}(\omega)|^{2}+|f_{1}(\omega)|^{2}(N-1)\big]\\
 & \times\big[\overline{n}(\omega)e^{i\omega\tau}+(\overline{n}(\omega)+1)e^{-i\omega\tau}\big]
\end{aligned}
$} & $\begin{aligned} & \langle\hat{D}_{S}^{(n)}(\tau)\hat{D}_{S}^{(m)}(0)\rangle=\int_{0}^{\infty}d\omega\mathcal{D}(\omega)\big[-2(g_{S}(\omega)-f_{0}(\omega))f_{1}(\omega)\\
 & +(N-2)|f_{1}(\omega)|^{2}\big]\times\big[\overline{n}(\omega)e^{i\omega\tau}+(\overline{n}(\omega)+1)e^{-i\omega\tau}\big]
\end{aligned}
$\tabularnewline
\hline 
\multicolumn{4}{|c|}{$\begin{aligned} & \langle\hat{D}_{S}^{(n)}(\tau)\hat{D}_{T}^{(n)}(0)\rangle=\int_{0}^{\infty}d\omega\mathcal{D}(\omega)(g_{S}(\omega)-f_{0}(\omega))(g_{T}(\omega)-f_{T}(\omega))\\
 & \times\big[\overline{n}(\omega)e^{i\omega\tau}+(\overline{n}(\omega)+1)e^{-i\omega\tau}\big]
\end{aligned}
$} & $\begin{aligned}\langle\hat{D}_{S}^{(n)}(\tau)\hat{D}_{T}^{(m)}(0)\rangle & =-\int_{0}^{\infty}d\omega\mathcal{D}(\omega)f_{1}(\omega)(g_{T}(\omega)-f_{T}(\omega))\\
 & \times\big[\overline{n}(\omega)e^{i\omega\tau}+(\overline{n}(\omega)+1)e^{-i\omega\tau}\big]
\end{aligned}
$ \tabularnewline
\hline 
\multicolumn{4}{|c|}{$\begin{aligned} & \langle\hat{D}_{T}^{(n)}(\tau)\hat{D}_{T}^{(n)}(0)\rangle=\int_{0}^{\infty}d\omega\mathcal{D}(\omega)|g_{T}(\omega)-f_{T}(\omega)|^{2}\\
 & \times\big[\overline{n}(\omega)e^{i\omega\tau}+(\overline{n}(\omega)+1)e^{-i\omega\tau}\big]
\end{aligned}
$} & \tabularnewline
\hline 
\end{tabular}\caption{\emph{Relevant correlation functions.} For simplicity in the notation,
we introduce $Q(A(\omega),\tau)=\int_{0}^{\infty}d\omega\mathcal{D}(\omega)\frac{A(\omega)}{\omega^{2}}\left[i\sin(\omega\tau)-\cos(\omega\tau)\coth(\beta\omega/2)\right]$,
$\Delta_{ST}^{(i)}=f_{i}(\omega)-f_{T}(\omega)$, and $\Delta_{S-ph}^{(i)}=f_{i}(\omega)-h(\omega)$
where, as defined above, $\mathcal{D}(\omega)$ is the vibrational
DOS. We also introduce the scaling factor $\lambda=\pm1$. \label{tab:correlation}}
\end{table*}
\end{widetext}

%\end{figure*}
%%%%

Once the time-domain correlation functions are calculated, we Fourier
transform them to obtain the rates of population transfer between
the chemical species of interest in our model. Given the variationally
optimized $\vec{X}(\omega)$, these rates can be analytically calculated
for the linear-linear and linear-exponential type correlations, by
using relations of the form

\begin{widetext}

\begin{align}
\int_{-\infty}^{\infty}\langle\hat{D}_{S}^{(n)}(\tau)\hat{D}_{S}^{(n)}(0)\rangle e^{i\omega_{0}\tau}d\tau & =\big[(g_{S}(-\omega_{0})-f_{0}(-\omega_{0}))^{2}+|f_{1}(-\omega_{0})|^{2}(N-1)\big]\overline{n}(-\omega_{0})\Theta(-\omega_{0})\nonumber \\
 & \hspace{1em}+\big[(g_{S}(\omega_{0})-f_{0}(\omega_{0}))^{2}+|f_{1}(\omega_{0})|^{2}(N-1)\big]\big[\overline{n}(\omega_{0})+1\big]\Theta(\omega_{0}),\label{eq:linear-linear}
\end{align}

\end{widetext}\noindent where $\bar{n}(\omega)=(e^{\beta\omega}-1)^{-1}$
is the boson occupation number and $\Theta(\omega)=1$ if $\omega>0$
and 0 otherwise. Eq. (\ref{eq:linear-linear}) can be readily understood
as a Fermi golden rule population transfer rate due to a single phonon
absorption (first row) or emission process (second row); expressions
like this also arise in standard Redfield theory \cite{Nitzan2006}.

On the other hand, the Fourier transform of the exponential-exponential
type correlation function is of the form $I=\int_{-\infty}^{\infty}e^{Q(A(\omega),\tau)}e^{i\omega_{0}\tau}d\tau$.
This integral can be approximated by means of the steepest descent
method \cite{Weiss2012}, which consists on the analytical continuation
of $Q(A(\omega),\tau)\rightarrow Q(A(\omega),z=\tau_{R}+i\tau_{I})$
(see Table \ref{tab:correlation}),

\begin{widetext}

\begin{align}
Q(A(\omega),z) & =-\int_{0}^{\infty}d\omega\mathcal{D}(\omega)\frac{A(\omega)}{\omega^{2}}\left[\coth(\beta\omega/2)\cos(\omega\tau_{R})\cosh(\omega\tau_{I})+\cos(\omega\tau_{R})\sinh(\omega\tau_{I})\right]\nonumber \\
 & \hspace{1em}+i\int_{0}^{\infty}d\omega\mathcal{D}(\omega)\frac{A(\omega)}{\omega^{2}}\left[\coth(\beta\omega/2)\sin(\omega\tau_{R})\sinh(\omega\tau_{I})+\sin(\omega\tau_{R})\cosh(\omega\tau_{I})\right].\label{eq:analytical_cont}
\end{align}

\end{widetext}

Let us rewrite $I=\int_{-\infty}^{\infty}e^{R(A(\omega),\tau)}d\tau$,
where $R(A(\omega),z)=Q(A(\omega),z)+i\omega_{0}z$, and find the
saddle point $z^{*}$ around which we will approximate the integral,

\begin{widetext}

\begin{equation}
\frac{dR(A(\omega),z^{*})}{dz}=\frac{\partial R(A(\omega),z^{*})}{i\partial\tau_{I}}=\frac{\partial\Im Q(A(\omega),z^{*})}{\partial\tau_{I}}+i\Bigg(\omega_{0}-\frac{\partial\Re Q(A(\omega),z^{*})}{\partial\tau_{I}}\Bigg)=0.\label{eq:R_min}
\end{equation}

\end{widetext}\noindent In Eq. (\ref{eq:R_min}), we took advantage
of the analyticity of $R(A(\omega),z)$, which guarantees the derivative
being independent of the direction of differentiation in the complex
plane. From Eqs. (\ref{eq:analytical_cont}) and (\ref{eq:R_min}),
we have that $\tau_{R}=0$. Therefore, we obtain
\begin{align}
\frac{dR(A(\omega),z^{*})}{dz} & =\int_{0}^{\infty}d\omega\mathcal{D}(\omega)\frac{A(\omega)}{\omega}\big[\coth(\beta\omega/2)\sinh(\omega\tau_{I}^{*})\label{eq:saddle_point_eq}\\
 & +\cosh(\omega\tau_{I}^{*})\big]+\omega_{0}=0,\nonumber 
\end{align}
where we notice that for $\omega_{0}=0$, the saddle point is at $\tau_{I}^{*}=-\frac{\beta}{2}$.
For $\omega_{0}\neq0$, we rely on the numerical solution of Eq. (\ref{eq:saddle_point_eq})
to find the root $z^{*}=i\tau_{I}^{*}$ such that

\begin{equation}
\int_{-\infty}^{\infty}e^{Q(A(\omega),z)}e^{i\omega_{0}z}dz\approx\left(\frac{1}{2\pi R''(A(\omega),z^{*})}\right)^{1/2}e^{R(A(\omega),\tau_{I}^{*})},\label{eq:Marcus_type_rate}
\end{equation}
where the prime denotes differentiation with respect to $\tau_{I}^{*}$,
and 
\begin{align}
R''(A(\omega),z^{*}) & =\int_{0}^{\infty}d\omega\mathcal{D}(\omega)A(\omega)\big[\coth(\beta\omega/2)\cosh(\omega\tau_{I}^{*})\label{second_deriv}\\
 & +\sinh(\omega\tau_{I}^{*})\big].\nonumber 
\end{align}

Eq. (\ref{eq:Marcus_type_rate}) is a generalization of the classical
Marcus rate, where $e^{R(A(\omega),\tau_{I}^{*})}$ is a Boltzmann
factor corresponding to an effective activation energy, and $\left(\frac{1}{2\pi R''(A(\omega),z^{*})}\right)^{1/2}$
is a density of states prefactor.

\section{Acknowledgments}

L.A.M.M is grateful for support of a UC-Mexus CONACyT scholarship
for doctoral studies. L.A.M.M. and J.Y.Z. acknowledge support of NSF
EAGER CHE-1836599. S.K.C. acknowledges support from the Canada Research
Chairs program and NSERC RGPIN-2014-06129. L.A.M.M. acknowledges Raphael
F. Ribeiro, Jorge Campos-González-Angulo, and Matthew Du for useful
discussions.

\bibliography{Polaron_photon_manuscript_jyz_w_corrections_031419}

\end{document}